\documentstyle[prl,aps,epsfig]{revtex}

\newcommand \beq{\begin{eqnarray}}
\newcommand \eeq{\end{eqnarray}}
\newcommand \bea{\begin{eqnarray}}
\newcommand \eea{\end{eqnarray}}

\newcommand \la{\raisebox{-.5ex}{$\stackrel{<}{\sim}$}}
\def\simge{\mathrel{%
       \rlap{\raise 0.511ex \hbox{$>$}}{\lower 0.511ex \hbox{$\sim$}}}}
\def\simle{\mathrel{
       \rlap{\raise 0.511ex \hbox{$<$}}{\lower 0.511ex \hbox{$\sim$}}}}
\def\simle{\mathrel{
       \rlap{\raise 0.511ex \hbox{$<$}}{\lower 0.511ex \hbox{$\sim$}}}}

\begin{document}
\title{Bose-Einstein transition in a dilute interacting gas.}
\author{Gordon Baym,$^{a,b,c}$
Jean-Paul Blaizot,$^b$
Markus Holzmann,$^{a,c}$
Franck Lalo\"e,$^{c}$ and
Dominique Vautherin$^d$\footnote{Our friend Dominique passed away on
December 6,
2000, before this manuscript was completed.}}
\address{$^a$University of Illinois at Urbana-Champaign,
1110 W. Green St., Urbana, IL 61801, USA}
\address{$^b$CEA-Saclay, Service de Physique Th\'{e}orique,
91191 Gif-sur-Yvette, Cedex, France}
\address{$^c$LKB and LPS, Ecole Normale Sup\'{e}rieure, 24 r. Lhomond,
75005 Paris, France}
\address{$^d$LPNHE, Case 200, Universit\'{e}s Paris 6/7, 4 Place Jussieu, 75005
Paris, France}

\maketitle
\begin{abstract}
       We study the effects of repulsive interactions on the critical 
density for
the Bose-Einstein transition in a homogeneous dilute gas of bosons.  First, we
point out that the simple mean field approximation produces no change in the
critical density, or critical temperature, 
and discuss the inadequacies of
various contradictory results in the literature.  Then, both within the
frameworks of Ursell operators and of Green's functions, we derive
self-consistent equations that include correlations in the system
and predict the change of the critical density.  We argue that the dominant
contribution to this change can be obtained within classical field theory and
show that the lowest order correction introduced by interactions is linear in
the scattering length, $a$, with a positive coefficient.  Finally, we
calculate this coefficient within various approximations, and compare with
various recent numerical estimates.
\end{abstract}

\section{Introduction}

        A precise description of the role of interparticle correlations on the
Bose-Einstein transition is indispensable to understanding its physical
nature; indeed, correlations are expected to play an essential role in the
very existence of superfluidity and related properties, vortices, flow
metastability, etc.  In general, dilute systems offer the possibility of
accurate microscopic treatments. The study of the  Bose-Einstein
transition in very dilute gases could provide experimental tests of the theory.
A large portion of the literature on the modification of the transition
temperature is based on a simple transposition of one of the most popular
method of condensed matter physics, mean field theory in various
guises, where the correlations are unmodified by the interactions and remain
purely statistical (as in an ideal gas).  Mean field theories,
for instance Gross-Pitaevskii, successfully describe a broad
variety of interesting phenomena observable in experiments, for example the
spatial distribution of the gas in a harmonic trap \cite{BP,HN}; for a recent
review of numerous successful applications of mean field theories in
Bose-Einstein condensation in atomic gases, see \cite {RMP}.  
Our purpose in this paper is to go beyond mean field theories and to
explore the effects of correlations on the properties of the
transition, studying in particular how they modify the transition temperature.

       We assume that the interparticle interaction can be
described by a positive scattering length $a$, equivalent to
the interaction of hard spheres of diameter $a$.  We shall also consider a
dilute gas, i.e., work in the regime where $a$ is much smaller than the
interparticle distance, $an^{1/3}\ll 1$, where $n$ is the particle density.
The critical number density, $n_{c}^0$, of an ideal gas is given by
\begin{equation}
n_{c}^0\lambda ^{3}=\zeta (3/2)\simeq 2.612,  \label{i1}
\end{equation}
where $\lambda $ is the thermal wavelength
\begin{equation}
\lambda =\frac{h}{\sqrt{2\pi mk_{B}T}},  \label{i2}
\end{equation}
$\zeta $ the Riemann zeta function; $m$ the particle mass and $k_B$
Boltzmann's constant.  Note that since at the transition $\lambda\sim
n^{-1/3}$, the diluteness condition is equivalent to $a\ll \lambda$.

       In an interacting gas, the critical value of the degeneracy parameter,
$n_c\lambda^{3}$, is modified; the first order change in the critical
temperature is related to that in the degeneracy parameter
by
\begin{equation}
      \frac{\Delta T_{c}}{T_{c}^{0}}
=-\frac{2}{3}
\frac{\Delta (n_{c}\lambda ^{3})}{n_{c}^0\lambda ^{3}}.
\label{i3a}
\end{equation}
Because mean field theories effectively treat physical systems as ideal
gases with modified parameters, the dimensionless degeneracy parameter keeps
exactly the same value as in an ideal gas.\footnote{ Mean field theories can
lead to a change of the effective mass \cite{FW}, which in turn affects the
value of the thermal wavelength in (\ref{i1}); with this effect included, the
critical value of the degeneracy parameter remains the same as for the ideal
gas.}  To calculate its change as a function of the
interactions it is necessary to go beyond mean field theories and include
correlations arising from the interactions.

        One might expect repulsive interactions to increase the degeneracy
parameter -- equivalently, to decrease the critical temperature, $T_c$, at
constant density;  in general, the presence of hard cores tends to impede the
motion of the particles necessary for quantum exchange effects, therefore
reducing the influence of quantum statistics and the critical temperature.
For example, the superfluid transition temperature of liquid $^4$He is below
that of an ideal gas of the same density.  Moreover, applying pressure to
liquid $^4$He effectively increases the role of the repulsive core of the
potential, and decreases the critical temperature.\footnote{In liquid $^3$He,
the repulsive cores similarly reduce the effect of quantum statistics, so that
the magnetic susceptibility is significantly higher than in an ideal Fermi
system with the same density.}

       Studies on the effect of interactions on the transition began in the 1950's
with the work of Huang, Yang, and Luttinger \cite{Huang1}, who concluded that
the phase transition of the interacting Bose gas ``more closely resembles an
ordinary gas-liquid transition than the Bose-Einstein condensation," but they
did not make a specific prediction for the change in $T_c$.  Shortly thereafter
Lee and Yang \cite{LeeYang57} predicted an {\it increase} of $T_c$
proportional to $a^{1/2}$; later, in Ref.
\cite{LeeYang58} they corrected this result and concluded
that the shift of the critical temperature is linear in $a$, with no
prediction for the magnitude or even the sign of the effect.
In 1960, Glassgold et al. predicted again a positive temperature shift
proportional to $a^{1/2}$ \cite{Glassgold}.
Later, Huang predicted an increase $\sim a^{3/2}$
\cite{Huang3}, and recently \cite{Huang-PRL}, he predicted that $T_c$
increases as $a^{1/2}$, using the same virial expansion as that of Ref.  \cite{LeeYang57}.  Despite the lack of qualitative agreement among
these many solutions of the problem, these studies showed that the changes in
question were not merely due to excluded volume effects (proportional to the
cube of the hard core diameter $a$) but to more interesting quantum effects,
proportional to a smaller power of $a$.

       The problem lay dormant for two decades until it was revisited by Toyoda
\cite{Toyoda}, who studied the transition in the Bogoliubov approximation in
the condensed phase.  This work predicted a {\it decrease} of the critical
temperatures at constant density proportional to $a^{1/2}$.  As the sign
agreed with the measurements in liquid $^4$He, the question appeared settled.
Toyoda's result was reinforced by numerical Path-Integral Quantum Monte Carlo
calculations showing that the effect of interparticle repulsion was indeed to
decrease the critical temperature \cite{MC1,MC2}.  Nevertheless, at the time
of these calculations, the issue did not have the same experimental interest
as it has now, and it was not fully appreciated that these calculations were
limited to relatively high densities and did not explore the region of dilute
systems.

        With the prospect of experimental realization of Bose-Einstein
condensation in dilute gases, Stoof \cite{Stoof1,Stoof2} carried out many-body
and renormalization group analyses concentrating on the dilute regime.
Stoof's work contains interesting precursors to the present work, e.g., Ref.
\cite{Stoof1} predicts a linear positive shift in the critical temperature
about twice that of our estimate in \cite{3/2club}.  Reference \cite{Stoof2}
predicts more structure in the $a$ dependence of the effect, $\sim a \ln a$,
\cite{Stoof3}, qualitatively similar to the $a^2 \ln a$ we describe below
\cite{NEW}.

        A surprise came when the Monte Carlo calculations for hard sphere bosons
were extended to lower densities and showed, in addition to the depression of
$T_c$ at high densities, the existence of a low density regime where the
critical temperature is indeed increased by the interaction \cite{GCL}.  At
very low densities the shift of the critical temperature was found to
be
\begin{equation}\label{deltaTc}
\frac{\Delta T_c}{T_c^0} = c \, n^{1/3} a,
\end{equation}
with $c \simeq 0.34$, determined by a numerical extrapolation to the limit
$a \to 0$. However, a more recent explicit Monte Carlo calculation \cite{HK}
of the leading correction to the ideal gas behavior predicts a prefactor
$c \simeq 2.3$. One source of the discrepancy lies in the
non-analytic dependence
of $T_c$ on $a$, discussed below, which gives rise to non-linear corrections
at the densities where the Monte Carlo calculation of Ref. \cite{GCL} was
performed.

       In the past several years, the problem was attacked by analytic 
approaches
based on self-consistent non-linear equations derived both in the Ursell
operator formalism \cite{HGL}, and the Green's function formalism
\cite{3/2club}.  One finds in both approaches that the effect of repulsive
interactions is to decrease the degeneracy parameter, thus increasing the
critical temperature at constant density.  
       Moreover, Ref.  \cite{3/2club} proves the linearity
of $\Delta T_c$ in $a$.  This was done by observing that the dominant
contribution to the shift in the critical density can be calculated by
restricting the propagators to their zero Matsubara frequency sector, thereby
reducing the quantum many-body problem to a classical field theoretical
problem in three spatial dimensions.  An alternative proof of the linearity in
$a$ based on renormalization group arguments is presented in \cite{BigN}.
While Refs.~\cite{HGL,3/2club,BigN} all agree on the functional form, they do
not provide definitive quantitative predictions for the prefactor; Ref.
\cite{HGL} provides $c\sim 1$, and an estimate in Ref.  \cite{3/2club} of an
exact formula for $\Delta T_c/T_c^0$ predicts $c\sim 2-3$.  In the limit of a
large number $N$ of components \cite{BigN}, $c=8
\pi/3\zeta(3/2)^{4/3}\simeq2.33$; interestingly, this exact result for $N\to
\infty$ agrees with the numerical result $c = 2.33 \pm 0.25$ of Ref.
\cite{HK} for $N=2$.  The reduction of the problem to classical field theory
has been exploited in the recent calculations of the transition in classical
$\phi^4$ field theory on the lattice extrapolated to the continuum
\cite{svistonov,arnold}, which give $c\simeq 1.3$.

        The linearity in $a$ is a non-trivial, non-perturbative result.
Since the
interaction is itself linear in $a$, one might imagine deriving this result in
some form of simple pertubation theory.  However, the first order term in $a$,
for fixed density, vanishes identically, while all higher order terms have
infrared divergences.  Nonetheless, various authors have attempted to skirt
the infrared problems.  For example, Ref.  \cite{Schakel} unjustifiably
``regularizes'' divergences in sums that appear at the transition with
an analytic continuation of the Riemann zeta function.\footnote{Indeed, the
method of Ref.  \cite{Schakel} applied to the simplest case of the
non-interacting Bose gas implies that as $\mu$ goes to zero $\partial n
/\partial \mu = \lambda_T^3 \zeta(1/2)/T$, which is finite and negative, in
contradiction to the divergence of the compressibility of the ideal gas.}
   Similarly,
Ref.  \cite{Huang-PRL} uses a virial expansion, unjustified at the critical
point, as we discuss below.  In another approach, Ref.  \cite{Wilkens}
attempts to exploit differences in first order pertubation theory between the
canonical and grand-canonical ensembles in finite volume; this
pertubative approach necessarily fails in the thermodynamic limit, preventing
a direct determination of the critical temperature. In Ref.~\cite{erich}
finite-size-scaling is used to reconcile this approach with the
grand-canonical calculations. Reference \cite{souza} calculates $T_c$
with the help of an ``optimized linear delta expansion´´, which avoids
infrared divergencies and operates for any $N$; for $N=2$ the authors
find $c \sim 3.0$, but the validity of the method is difficult to assess,
and the accuracy of this result may be affected by uncontrolled errors.

       The aim of this paper is to summarize current understanding of 
the problem
of the transition temperature.  We provide a more detailed account of our
earlier analytical calculations, and in addition compare the Green's function and
Ursell calculations.  The paper is organized as follows:  In the next section
we recall features of the Bose-Einstein transition in an ideal gas and show
how the addition of a mean repulsive field does not alter the critical value
of the degeneracy parameter.  Then in Sec.~III we include correlations and
obtain, using alternatively the Ursell and Green's function formalisms, simple
self-consistent equations which reveal the physical origin of the change in
the critical temperature.  In Sec.~IV we show that the dominant contribution
to the change in the critical temperature can be calculated using a classical
field approximation, and we show that the resulting change is linear in the
scattering length.  Section V is devoted to numerical calculations of the
coefficient $c$, and to a numerical exploration of the range of validity of
the linear behavior. We focus throughout on a spatially uniform system; a
discussion of the transition temperature of a dilute gas in a
trap can be found in Refs.~\cite{HN,stooftrap,arnoldtrap}.
For experimental data in the $^4$He-Vycor system, see \cite{reppy}.

\section{Ideal gas, mean field and related calculations}

\label{ideal}

        In a homogeneous system, the number density of a non-condensed 
ideal Bose
gas is given by
\beq
n= \int \frac{d^3k}{(2\pi)^3}
\frac{1}{e^{\beta (\varepsilon_k^0 -\mu)}-1}
=\frac{1}{\lambda^3} g_{3/2}(z)
\label{ig1}
\eeq
with $\beta =1/k_{B}T$,
$\mu $ is the chemical potential, $z=\exp(\beta \mu)$ the fugacity,
and ($\hbar=1$)
\beq
\varepsilon_k^0=k^2/2m ;
\eeq
the Bose (polylogarithmic) function $g_{p}(z)$ is defined by
\begin{equation}
g_{p}(z)\equiv \sum_{j=1}^{\infty }\frac{z^{j}}{j^{p}}.  \label{ig2}
\end{equation}
As $\mu $ tends to zero from negative values $g_{3/2}(z) \to
\zeta (3/2)$ corresponding to the
maximum density for a non-condensed gas at a given temperature
given by Eq. (\ref{i1}).

        The simplest way to include repulsive interactions is  in
mean field.
Assuming that all the effects of interactions can be described
by an $s$-wave scattering length, one can generalize Eq.~(\ref{ig1})
by writing:
\begin{equation}
n=\frac{1}{\lambda ^{3}}g_{3/2}(e^{\beta (\mu -\Delta \mu )})  \label{ig3}
\end{equation}
where the shift of the chemical potential $\Delta \mu $ is proportional to
the number density:
\begin{equation}
\beta \Delta \mu = 2g n \beta=4a\lambda ^{2}n,  \label{ig4}
\end{equation}
and $g=4 \pi \hbar^2a/m$.  Equation (\ref{ig3}) is a simple consequence
of the Hartree-Fock approximation, using a pseudopotential proportional to
$a$, in which the shift of the single particle energies is given by
\begin{equation}
\Sigma _{HF}=2gn;  \label{ig5}
\end{equation}
the factor of two comes from exchange. Since $\Sigma_{HF}$ is independent of
momentum we have to increase the chemical potential by $\Delta
\mu=\Sigma_{HF}$ to keep the same particle density as the ideal gas. The same
results are obtained in \S 4 of Ref. \cite{HGL}.

\begin{figure}
\begin{center}
\epsfig{file=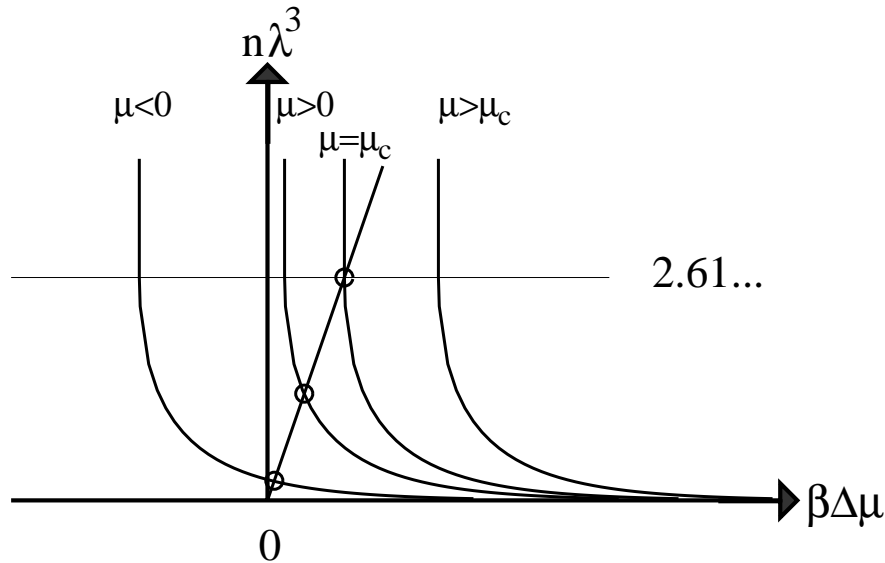,height=8cm}
\end{center}
\caption{\label{figmu3}
Plot of the density $n\lambda^3$ as a function of the shift of
the chemical potential $\beta \Delta \mu $, for different values of the
chemical potential $\beta \mu $. For a given
value of $\beta \mu$, the self-consistent solution is given by the
intersection point (circle) of this curve with the straight line $\beta
\Delta \mu =4a\lambda^2 n$.
}
\end{figure}

\begin{figure}
\begin{center}
\epsfig{file=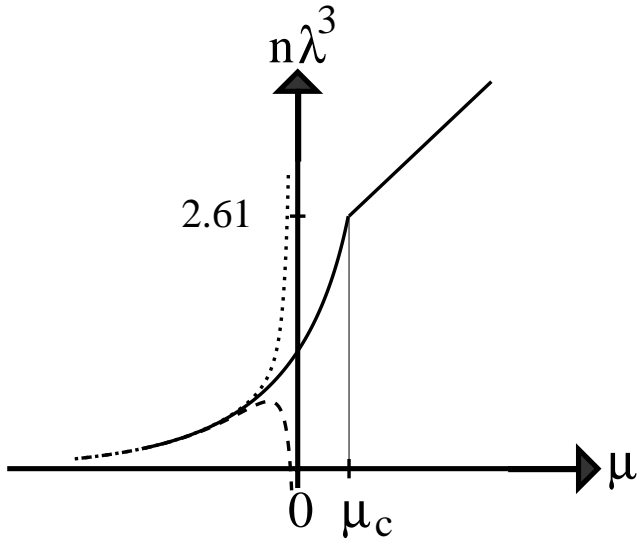,height=8cm}
\end{center}
\caption{\label{figdens5}
Variations of the density $n\lambda^3$ as a function of the
chemical potential predicted in mean field (full line), for
small constant
value of the interaction parameter $a/\lambda $. The dotted line
corresponds to the
ideal gas. The mean field solution curve shows no maximum, but a constantly
increasing density as a function of the chemical potential. On the
other hand, any
finite expansion in orders of
$a/\lambda $ leads to divergencies around $\mu =0$ and may introduce
spurious extrema
(dashed line).  }
\end{figure}

       Because the Hartree-Fock self-energy depends on the density, the relation
between the chemical potential at the transition and the critical density is
more complicated than in the non-interacting case, and the equation
$\mu-\Delta\mu=0$ is non-linear.  Its solution is
conveniently obtained with the geometrical method of \cite{HGL}, illustrated
in Fig.~\ref{figmu3}.  At fixed $\mu $ and $\beta$, with $\beta
\Delta \mu $ as a
variable, the density is obtained through (\ref{ig3}); then a simple
construction provides the value of $ \Delta \mu $ corresponding to the
transition.  Finally, the density as a function of $\mu $ varies as shown in
Fig.~\ref{figdens5} (full line); it behaves similarly to that of
the ideal gas.  However,
the transition now occurs at a positive value of the chemical potential and
the compressibility $(1/n^2)\partial n/\partial \mu$ is finite, in contrast to
the ideal gas where it diverges.  As mentioned in the introduction, the
critical density remains exactly the same as for the ideal gas, because  
at the transition $\mu =\Delta \mu $ and thus critical density is given
by the same integral as for the ideal gas.

       It is instructive to use this simple mean field model to test 
the limit of
simple approximations in an exactly soluble case.  Expanding the right side of
Eq.  (\ref{ig3}) in powers of $a/\lambda $ we find the virial expansion:
\begin{equation}
n\lambda ^{3}=g_{3/2}(z)-4\frac{a}{\lambda }g_{1/2}(z)\times
g_{3/2}(z)+8\left( \frac{a}{\lambda }\right) ^{2}g_{3/2}(z)\left\{ 2\left[
g_{1/2}(z)\right] ^{2}+g_{3/2}(z)g_{-1/2}(z)\right\} + \dots.
\label{ig6}
\end{equation}
Now, if as Ref.~\cite{Huang-PRL} we consider only the first two terms of the
expansion we find that the density (the broken lines in
Fig.~\ref{figdens5}) develops a
maximum for a negative value of $\mu $; since the density must always
be an increasing function of $\mu $, it then becomes tempting to infer that a
phase transition should take place at this point.  As seen in the figure, this
point corresponds to a smaller density than for the ideal gas; this reasoning
would then predict an increase of the critical temperature, at constant
density, proportional to $\sqrt{a}$, precisely the result obtained in
\cite{Huang-PRL}.  But one should keep in mind that in this simple model the
density maximum is just an artefact of the first order virial expansion, as
illustrated by the absence of any maximum in the full curve of
Fig.~\ref{figdens5}; in
fact, inclusion of second order terms in $a$ of (\ref{ig6}) 
makes the maximum disappear.\footnote{The second order terms given in Ref.
\cite{Huang-PRL} differ from those of 
Eq.~(\ref{ig6}); nevertheless they do not change our argument.} 
From the original  Eq.~(\ref{ig2}), the critical density cannot change. 
Similar
arguments were already given in Refs.~\cite{FL} and
\cite{HGL}; sufficiently close to $ \mu =0$, higher order terms diverge
faster and, eventually, dominate the lower order terms in any viral
expansion.\footnote{The end of the discussion of \S 3.3 of Ref.  \cite{FL} was
given specifically for the case of attractive interactions; then, instead of a
density maximum, the naive first order virial correction model predicts the
disappearance of the transition, which is replaced by a simple crossover
between two regimes.  For repulsive interactions, the sign of the first order
correction is opposite, and the density maximum occurs as in Ref.
\cite{Huang-PRL}.}

       The simple example above illustrates the dangers of
truncating an expansion in $a$, even within the mean field approximation.  The
physical origin of the difficulty is simple:  the very essence of
Bose-Einstein condensation is the appearance of long exchange cycles over the
system, which cluster together all particles that they contain \cite{Elser};
therefore, the phenomenon is not easily captured within any formalism
containing a limitation on the size of clusters; further discussion of the
effect of long exchange clusters on the position of the Bose-Einstein
transition can be found in \cite{GCL}.  One must be very careful in truncating
pertubative expansions in which nominally higher-order terms turn out to be of
comparable order; rather it is necessary in general to sum an infinite number
of terms.

        In the calculation of Toyoda \cite{Toyoda}, the only mean field
taken into
account is that due to the condensed particles below the transition
temperature.  His approximation is in fact lowest order Bogoliubov theory.
While this theory describes correctly the ground state at zero temperature and
its elementary excitations, its extension near the critical temperature meets
several difficulties; in particular it predicts a first order phase transition
\cite{BG}, a point not taken into account.  In fact, above $T_c$, Toyoda's
calculation of the free energy is just that of an ideal gas, with no shift in
the critical temperature.

\section{Self-consistent equations}

\label{non-linear-equations}

       In this section, as well as in the rest of this paper, we concentrate on
the non-condensed state and approach the critical temperature from above.  As
we have seen in the previous section, mean field effects which produce merely
a constant shift in the single particle energies around $T_{c}$ do not affect
the value of the critical temperature.  A modification of $T_{c}$ thus
requires the inclusion of correlations; whose effect of such
correlations is to lower the single particle occupation at small $k$.  Thus, as
the temperature is lowered, the chemical potential  reaches the lowest single particle
energy
for a value smaller than in mean field, resulting in a decrease of
$n_{c}$.

       It is instructive at this stage to consider the highly oversimplified
model in which correlations push down only the level $k=0$ and all other
levels are treated within mean field.  Then, at the transition, when the $k=0$
level hits the chemical potential, the other particles experience a constant
energy shift $\bar\mu>0$ with respect to the level $k=0$, and  the critical
density is given by:
\begin{equation}
n_{c}=\int \frac{d^3k}{(2\pi)^3}\frac{1}{e^{\beta(\varepsilon_k^0 +\bar\mu)}
-1}=\frac{1}{\lambda^3}\,g_{3/2}\left( e^{-\beta \bar\mu} \right).
\label{d3}
\end{equation}
Since $\bar\mu$ is small, one can expand the Bose function
  $g_{3/2} \left( e^{-\beta \bar{\mu}} \right)$:
\begin{equation}
g_{3/2}\left( e^{-\beta \bar{\mu}} \right)=
\zeta(3/2) -2 \sqrt{\pi \beta \bar{\mu}}.
\label{d4}
\end{equation}
In fact, in calculating
the change in the critical density $\Delta n_c=n_c-n_c^0$, one can
equivalently expand the statistical factor in (\ref{d3}) at small
$k$:
\begin{equation}
\frac{1}{e^{\beta(\varepsilon_k^0 +\bar\mu)}-1}\to
\frac{1}{\beta(\varepsilon_k^0
+\bar\mu)}
\label{d5}
\end{equation}
and arrive at the result:
\begin{equation}
\Delta n_c=n_c-n_c^0\approx \int_0^\infty
\frac{dk}{\pi^2}\frac{m}{\beta} k^2\left(
\frac{1}{k^2+2m\bar\mu}-\frac{1}{k^2}\right)=
-\frac{2}{\lambda^3}\sqrt{\pi\beta\bar\mu}.
\label{d6}
\end{equation}
We shall often use the approximation (\ref{d5}) in the following.  We note
here that it is valid provided only momenta $k\ll \lambda^{-1}$ contribute
significantly to the integral (\ref{d6}), which requires $\sqrt{2m\bar\mu}\ll
\lambda^{-1} $. Since we expect the first correction beyond mean field to be
$\beta\bar\mu \sim (a/\lambda)^{2}$, this condition is satisfied if
$a\ll \lambda$.

       As anticipated, the correlations that push down the level
$k=0$ lead to a decrease of the critical density, and hence to an increase of
the critical temperature.  Furthermore, the magnitude of the
effect is not necessarily analytic in the small change $\bar\mu$
of the chemical potential, and hence in the interaction strength $a$.  In
fact, the expected result $\beta\bar\mu \sim (a/\lambda)^{2}$ together with Eq.
(\ref{d6}) lead to $\Delta n_{c}/n_c \sim a/\lambda$.
       To include correlations more generally we consider two 
approaches, that of
Ursell operators used already for this problem in \cite{HGL}, and
that of finite
temperature field theory.  Before deriving detailed results, let us
spend a moment
comparing the two approaches.

Within finite temperature field theory one
typically carries out a systematic expansion of the properties of a
many-body system,
e.g., the pressure, in powers of the interaction
$V$, using either
unperturbed Green's functions $G_0$ or self-consistent ones, $G$.
Ursell operators provide a different approach to calculating 
thermodynamic
properties of an interacting many-particle system, and lead naturally to
expansions in terms of correlations of higher and higher orders.  The
Ursell operator of
rank
$n$,
$U_n$, describes the correlations of a system of $n$ interacting
Boltzmann particles.
For example, the operator $U_2$, defined by
\beq
U_2=e^{-\beta \left( p_1^2/2m +p_2^2/2m +V(r_1-r_2) \right)}
- e^{-\beta ( p_1^2/2m +p_2^2/2m)},
\eeq
acounts for two-body correlations.  One expects that matrix
elements of $U_n$ vanish between states in which  one of the particles is
far away from the others, and, in the tradition of cluster expansions,
one writes expansions of thermodynamic functions in powers of the
Ursell operators $U_n$.  Every term of such an expansions is expected
to be finite, even for highly singular potentials such as hard spheres.
Inclusion of the specific bosonic or fermionic statistics gives rise to
exchange cycles.

       Solved exactly, both formalisms give in principle identical results for
static thermodynamic properties, and detailed comparisons of how specific
approximations can be formulated in either approach can be found in
\cite{Markus}.  In the following we  study the effects of
correlations by means of
a simple self-consistent approximation which can be derived in either
formalism.
This simple self-consistent approximation leads to a
nonanalytic change
in the spectrum at small $k$.

\subsection{Ursell operators}

\label{Ursell op}

        We briefly summarize the principal results obtained in
\cite{HGL} with the
Ursell method, reformulated here in a way to make a ready comparision with the
Green's function approach. More details are given in Appendix.

        Reference \cite{HGL} provides the general diagramatic rules to 
obtain the
     reduced one-body density operator in momentum space,
$\rho_k$, as a function of the ideal gas Bose distribution, $f_k$, and the
Ursell operators $U_n$ ($n \ge2$).  Quite generally, above the
critical point the single particle density operator has the form of
a Bose distribution, but with modified single particle energies,\footnote{Since
$\delta
\mu_k$ is real and plays the role of correcting the ideal gas energy
in the Bose
distribution, the energies
$k^2/2m+\delta \mu_k$ may be regarded as those of statistical quasiparticles,
in the sense of Ref.~\cite{Balian} for the Fermi liquid. Such
statistical quasiparticles are not equivalent to those obtained from the
Green's functions.}
\beq
\rho_k =f_k(\mu - \delta \mu_k)
=\frac{1}{e^{\beta (\varepsilon_k^0 +\delta \mu_k -\mu)}-1}.
\eeq
      The particle
number density is given in terms of $\rho_k$ by
\beq\label{densityrhok}
n=\int \frac{d^3 k}{\left( 2\pi \right) ^{3}} \,\rho_k.
\eeq

When looking for leading order corrections one can safely ignore 
Ursell operators, $U_n$, with $n \ge 3$.  The resulting
topological structure of the diagrams
of the Ursell pertubation series then becomes equivalent to that of
the perturbative
expansion using Green's functions with two-body interactions.  Furthermore,
by treating
the matrix elements of
$U_2$ as momentum independent, one quantitatively recovers the
pertubation theory of
the Green's function approach with a momentum-independent coupling
constant related to
the
$s$-wave scattering length $a$.  Finally, as we shall below, at the
critical point the density operator $\rho_k$ becomes large at small momenta,
$\rho_k \gg 1$ for $k \to 0$, so that the approximation
\beq
1+\rho_k \simeq \rho_k \simeq [\beta (\varepsilon_k^0+\delta \mu_k -\mu)]^{-1}
\eeq
can be used systematically (as in (\ref{u10}) below). These remarks
explain why the
results that we obtain using Ursell operators will eventually
be identical to
those obtained within the Green's function approach in the particular
limit of $a \ll \lambda$.

\begin{figure}
\begin{center}
\epsfig{file=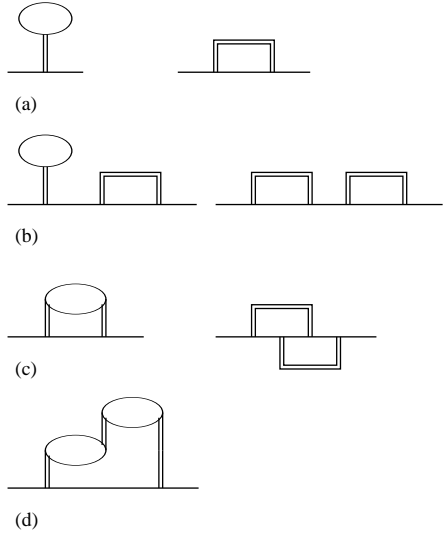,height=8cm}
\end{center}
\caption{\label{ursellrho}
(a) First order Ursell diagrams; the double line corresponds to
one operator $U_2$. Here we slightly change the representation of
Ref.~[21] and
replace the dotted segments there (corresponding to a summation over all
exchange cycle
lengths) by closed curves; in this way, the $U_2$ diagrams become very
similar to the
usual Green's function diagrams. Despite this close graphical
similarity, the physical
interpretation of the diagrams is different: for instance, exchange
cycles do not
appear at all in Green's function diagrams. (b) Simplest examples of
the diagrams
included in the iteration. (c) Diagrams leading to Eqs.~(\ref{uu6})
and (\ref{u5ter}).
(d) Bubble diagrams leading to Eq.~(\ref{u9bis}).
}
\end{figure}

        To obtain the mean field result, Eq.~(32) of Ref.~\cite{HGL},
we iterate the first order diagrams shown in Fig.~\ref{ursellrho}a
(examples of iterated
diagrams are shown in Fig.~\ref{ursellrho}b); this leads to:
\begin{equation}
\rho_k=f_{k}(\mu -\Delta \mu ),  \label{u1}
\end{equation}
with
\begin{equation}
\beta \Delta \mu \simeq
4a\lambda^2
\int \frac{d^3 k }{(2\pi)^3 } \,\rho_k=4a\lambda^2 n.  \label{u2ter}
\end{equation}
In this approximation the only effect of the interactions is to
produce a momentum
independent shift of the single particle energies which, as discussed in the
previous section,  can be absorbed in a shift
of the chemical potential:
\begin{equation}
\mu^{\prime}= \mu -\Delta \mu,  \label{u4}
\end{equation}
leaving the critical density identical to that of the
ideal gas.

        To go beyond mean field, we include in the self-consistent equation for
$\rho$ the corrections displayed in Fig.~\ref{ursellrho}c. These are
formally of
second order in
$a/\lambda
$ and read (\S5 of \cite{HGL}):
\begin{equation}
\rho_k = f_k (\mu^{\prime} -\delta \mu_k),  \label{u5}
\end{equation}
with
\begin{equation}
\beta \delta \mu_k \simeq
-8\left( \frac{a}{\lambda } \right)^2
\lambda^6
\int \frac{d^3 k^{\prime}}{( 2\pi )^3}
\int \frac{d^3 q}{( 2\pi )^3}
\, \rho_{k^{\prime}}
\, \rho_{k^{\prime}-q }
\,\rho_{k+q}.
\label{uu6}
\end{equation}
Note how the integral (\ref{uu6}), in which
momentum conservation appears explicitly ($q$ is the momentum transfer in a
binary collision), introduces a $k$-dependence of the energy shift, as
opposed to the result of the simple mean field approximation.

        We assume that the single
particle state with $k=0$ still has the lowest energy, so
that the phase
transition occurs when
\begin{equation}
\mu^{\prime}
-\delta \mu_{k=0} =0.  \label{u7}
\end{equation}
The critical density is then given by
\begin{equation}
n_c=\int \frac{d^3 k}{( 2\pi )^3}
\,\rho _k=\int
\frac{d^3 k}{( 2\pi )^3}
\,  f_k(\delta \mu_{k=0}-\delta
\mu_k),  \label{u8}
\end{equation}
which, because of the $k$-dependence of $\delta \mu _{k}$, does not
coincide with the critical density of the ideal gas obtained for constant
$\delta \mu$.
Instead
\begin{equation}
\Delta n_c=n_c-n_c^0=\int
\frac{d^3 k}{( 2\pi )^3}
\, \left[ f_k(\delta \mu_{k=0}-\delta
\mu_k)-f_k(\delta \mu_{k=0})\right].\end{equation}
     In general, $\delta \mu _{k}-\delta \mu _{k=0}$ is an increasing
function of $k$,
so that
$\Delta n_c$ is negative.

        The variable appearing in Eq.~(\ref{u8}),
\begin{equation}
\delta \mu_k -\delta \mu_{k=0}
\simeq -\frac{8}{\beta }\left( \frac{a}{
\lambda }\right)^2 \lambda^6 \int \frac{d^3 k^{\prime}}{( 2\pi)^3}
\int \frac{d^3 q}{( 2\pi )^3} \,\rho_{k^{\prime}}
\,\rho _{k^{\prime}-q}\,\left[ \rho_{k+q}
-\rho_q \right],  \label{u8bis}
\end{equation}
can be simplified if we notice that when the
critical condition (\ref{u7}) is fulfilled the dominant contribution  to the
integrals comes from small momenta for which the statitical factors $f_k$
diverge.  In fact, if the
$f_k$'s are evaluated with the free particle spectrum, the integral
in (\ref{uu6})
becomes logarithmically divergent in the infrared.  To see that, we expand the
$f_{k}$'s as in (\ref{d5}), 
\begin{equation}
\left[ e^{\beta (\varepsilon_k^0-\mu^{\prime}+\delta \mu_k )} -1 \right]^{-1}
\simeq
\frac{1}{\beta ( \varepsilon_k^0+\delta \mu_k-\mu^{\prime})}.
\label{u10}
\end{equation}
Setting
\beq
\varepsilon_k=\varepsilon_k^0+\delta \mu_k-\delta \mu_{k=0},
\eeq
we obtain the self-consistent relation valid at small $k$,
\begin{equation}
\varepsilon_k=\varepsilon_k^0
-8\left( \frac{a\lambda ^2}{\beta^2}\right)^2
\int \frac{d^3 k^{\prime}}{( 2\pi )^3}
\int \frac{d^3 q}{(2\pi)^3}
\frac{1}{\varepsilon_{k^{\prime}}}
\frac{1}{\varepsilon_{k^{\prime}-q}}
\left[ \frac{1}{\varepsilon_{k+q}}
-\frac{1}{\varepsilon_q}\right].  \label{u11}
\end{equation}
    A simple power counting argument indicates the integral is logarithmically
divergent if we replace the self-consistent energies $\varepsilon$
by the  free
$\varepsilon_k^0$.  For the  self-consistent spectrum, however,
no infrared
divergences occur, as we shall see in ßIII C.

\subsection{Green's functions}

\label{Green}
In the
normal state the single particle Green's function $G(k,z_{\nu })$ is given
\beq
G^{-1}(k,z_{\nu })=z_{\nu }+\mu -\varepsilon_k^0-\Sigma(k,z_{\nu}),
\label{g2}
\eeq
where $k$ is the single particle momentum, and  $z_{\nu}=2\pi i \nu/\beta$
is a Matsubara frequency, with $\nu=0,\pm 1,\pm 2,\dots$.  The
self-energy $\Sigma(k,z)$, which describes the effect of the interactions,  can be
obtained as a series in powers of the interaction strength $a$ by
standard diagrammatic techniques
\cite{FW,BR,KB,self-consistent}. The single particle density matrix
is related to
$G$  by
\begin{equation}
\rho_{k}=-T\sum_{\nu
}\,\,e^{z_{\nu }0^+ }\,G(k,z_{\nu }).  \label{g1}
\end{equation}

        The criterion for condensation is  that the chemical potential
$\mu$ reaches
the bottom of the single particle excitation spectrum, and we again assume, as in
\S\ref{Ursell op}, that the lowest single particle state is that with
$k=0$. The
transition point is then determined by the condition
\cite{PPbook}:
\begin{equation}
G^{-1}(0,0)=0 \quad \text{or} \quad \Sigma
(0,0)=\mu \,.  \label{g3}
\end{equation}
        At that point,
\begin{equation}
G^{-1}(k,z_{\nu })=z_{\nu }-\varepsilon_k^0-\left[\Sigma(k,z_{\nu})
     -\Sigma(0,0)\right],  \label{g4}
\end{equation}

       To first order in the interaction strength, the self-energy is given by
the Hartree-Fock approximation,\footnote{ Implicit in this expression is the
summation of two-body collisions via the $t$-matrix, which relates the
two-body potential to the scattering length $a$ at low energies.
The summation is made with diagrams where the intermediate
propagators are free, corresponding to two particles interacting in the vacuum
\cite{Belyaev}.} leading to a contribution $\Sigma_{HF}$ (see Eq.~(\ref{ig5}))
independent of both
$k$ and $z$ which can then be eliminated by a redefinition of the chemical
potential, as discussed above.

\begin{figure}
\begin{center}
\epsfig{file=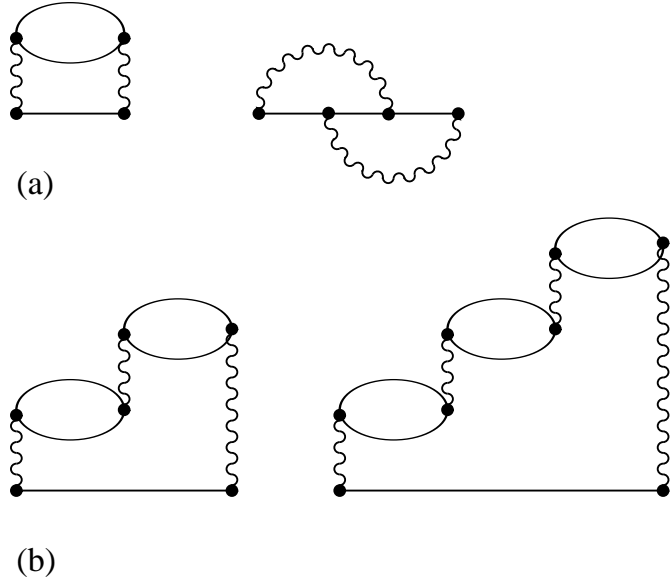,height=8cm}
\end{center}
\caption{\label{figgreen1}
(a) Green's function diagrams leading to Eq.~(\ref{g7}),
similar to Fig. 3c.
(b) Green's function bubble diagrams leading to Eq.~(\ref{f10}),
similar to Fig. 3d.}
\end{figure}

        The structure of $\Sigma (k,z)$ in next order is
described by the two diagrams in Fig. ~\ref{figgreen1}a.  The
second is
the exchange term of the first, and within the present approximation
in which the
matrix elements of the interaction do not depend on momenta, the two
contributions are equal.  Replacing the free propagators by their
Hartree-Fock version, we  have
\begin{equation}
\Sigma (k,z_\nu)=
2 g^2
\int \frac{d^3 k^{\prime}}{(2\pi)^3} \frac{d^3q}{(2\pi)^3}
\frac{ f_{k^{\prime}} (1+f_{k^{\prime}+q}) (1+f_{k-q})
- (1+f_{k^{\prime}})f_{k^{\prime}+q} f_{k-q}}{
z_\nu +\mu^{\prime}+\varepsilon_{k^{\prime}}^0-\varepsilon_{k^{\prime}+q}^0
-\varepsilon_{k-q}^0}.   \label{g7}
\end{equation}
Because the condensation condition (\ref{g3}) involves only the Matsubara
frequency $z_\nu=0$, we concentrate from now on this contribution.
Furthermore, as before, we isolate the dominant contribution
by expanding the statistical factors, so that
$f_{k}\sim T/\varepsilon_k^0$, and
\begin{equation}
\Sigma (k,0)-\Sigma (0,0)\simeq - 2g^2 T^2
\int \frac{d^3 k^\prime}{(2\pi)^3} \frac{d^3 q}{(2\pi)^3}
\frac{1}{(\varepsilon_{k^\prime}^0 - \mu^\prime )
(\varepsilon_{k^\prime-q}^0 - \mu^\prime)}
\left( \frac{1}{\varepsilon_{k-q}^0 - \mu^\prime}
- \frac{1}{\varepsilon_q^0-\mu^\prime} \right).
\label{g10}
\end{equation}
Replacing the bare energies $\varepsilon_k^0$
by the dressed energies $\varepsilon_k^0+\Sigma (k,0)$, observing that
$\mu'=\Sigma (0,0)$ at the transition, and setting
\begin{equation}
\varepsilon_k=\varepsilon_k^0+\Sigma (k,0)-\Sigma (0,0),  \label{g11}
\end{equation}
we recover Eq.~(\ref{u11}). Note that the earlier $\delta
\mu_{k}-\delta \mu_{k=0}$
is simply $\Sigma (k,0)-\Sigma (0,0)$.

\subsection{Discussion}

\label{discussion}

        The change of the critical density is intrinsically related to
$\delta \mu
_{k}$ (or equivalently $\Sigma(k,0)$) by Eq.~(\ref{u8}). In terms of
\beq
U(k)=2m \left[\delta \mu_{k}- \delta \mu_{k=0}\right]=2m \left[
\Sigma (k,0)-\Sigma
(0,0)\right],
\eeq
the change of the critical density introduced by the interaction is 
\beq
\Delta n_{c}  = \int \frac{d^3k}{(2\pi)^3} \left\{
\frac{1}{e^{\beta [k^2 + U(k)]/2m}-1} - \frac{1}{e^{\beta k^2/2m} -1}
\right\}     \label{firstl}
\eeq
or
\beq
\Delta n_c \simeq  -\frac{2}{\pi \lambda^2}
     \int_{0}^{\infty }dk\, \frac{U(k)}{k^2+U(k)},  \label{u13}
\eeq
where in the second line we have used the approximation (\ref{d5}).

The heart of the calculation of the
critical density is then to determine the function $U(k)$, a
non-trivial task, since the
evaluation of this function by naive perturbation expansion fails because of
infrared divergences. However, higher order iterations lead to
an instability in the energy spectrum at small momenta, as in 
Ref.~\cite{HGL}. In the limit 
of an infinite number of iterations, the spectrum around
$k=0$ hardens: the self-consistent solution
of (\ref{uu6}) leads indeed to $\varepsilon_{k}\sim k^{3/2}$, as predicted by
Patashinskii and Pokrovskii \cite{PP} using the following argument.

     For free
particles, the integral in Eq.~(\ref{g10}) contains six powers of momentum in
both the numerator and denominator, and is thus
logarithmically divergent. In order to ensure that the self-consistent
solution, $\varepsilon_k$, converges in the infrared limit, $\varepsilon_k$
must behave (modulo possible logarithmic corrections) as $\sim k^{\alpha }$
with $\alpha <2$, so that the free particle energies, $\varepsilon_k^0 \sim
k^2$, can be neglected at small $k$ with respect to $\Sigma(k,0)-\Sigma(0,0)$.
With this behavior, $\Sigma (k,0)-\Sigma (0,0)\sim k^{6-3\alpha }$, so that we
find a self-consistent energy spectrum, $\varepsilon_k \sim \Sigma
(k,0)-\Sigma (0,0)\sim k^{\alpha }$, for $\alpha =3/2$.

The modification of the spectrum occurs only for small momenta $k\ll
k_c$, where $k_c$ is
a scale that will be specified below. We note here only that since
$U(k)$ is of order
$a^2/\lambda^4$, one expects $k_c$ to be of order $a/\lambda^2$. For
momenta $k_c \ll k \to \infty$,
perturbation theory  becomes applicable leading to
$U(k)/k^2\to 0$. The typical momenta involved in the integral
(\ref{u13}) are of order
$k_c$. The validity of Eq.~(\ref{u13}) for $k\ll
\lambda^{-1}$ requires
$k_c\ll\lambda^{-1}$, which is satisfied in the dilute limit.

     We later present numerical self-consistent solutions of
Eq.~(\ref{u11}). Here we reconsider the simple analytical model
calculation of Ref.
\cite{3/2club} which provides an estimate for the scale
$k_c$, and acts as a reference for the numerical results presented later.
In this analytical model we construct a self-consistent energy
spectrum at the critical
point:
\begin{equation}
\varepsilon_k = \frac{\hbar^2 k^2}{2m} + \Sigma(k,0) -\Sigma(0,0)
\label{noname}
\end{equation}
within the approximation (\ref{g10}) for the self-energy, which we write as
\begin{equation}
\Sigma (k,0)-\Sigma (0,0)=-2g^{2}T\int \frac{d^3q}{(2\pi )^3} B(q) \left(
\frac{1}{\varepsilon_{k-q}}-\frac{1}{\varepsilon_q}\right),
\label{smallk0}
\end{equation}
where the bubble diagram contributes
\begin{equation}
B(q)=T\int \frac{d^{3}p}{(2\pi )^{3}}\frac{1}{\varepsilon_{p}
\varepsilon_{p+q}}.  \label{bubb0}
\end{equation}
To extract the low
momentum structure, below the scale $k_{c}$, we evaluate the most divergent
terms of Eq. (\ref{smallk0}) using the following ansatz:
\begin{equation}
\varepsilon_k=k_c^{1/2} \frac{\hbar^2 k^{3/2}}{2m}
\Theta(k_c-k) + \frac{\hbar^2 k^2}{2m} \Theta(k-k_c).
\label{ansatz32}
\end{equation}
With this spectrum,
Eq.~(\ref{bubb0}) becomes
\begin{equation}
B(q) \simeq \frac{4m}{\pi \hbar^2 \lambda ^{2}k_{c}}\left[ \ln
\left(\frac{k_{c}}{q} \right)+c\right],
\label{f4}
\end{equation}
where $c\approx 2+2\ln 2-\pi /2$ =1.816, and in the limit $k\to 0$ the self-energy is
\begin{equation}
\Sigma (k,0)-\Sigma (0,0)=\frac{1024\pi \hbar^2}{15m}\left( \frac{a}{\lambda ^{2}}
\right) ^{2}\left( \frac{k}{k_{c}}\right) ^{3/2} . \label{smallk}
\end{equation}
Identifying the right side of this equation with
$k_{c}^{1/2}\hbar^2 k^{3/2}/2m$, the self-consistency condition,
Eq.~(\ref{ansatz32}), in the limit $k \to 0$ applies that
\begin{equation}
k_{c}=32\left( \frac{2\pi }{15}\right) ^{1/2}\frac{a}{\lambda ^{2}}\approx
20.7\frac{a}{\lambda ^{2}}.  \label{kc}
\end{equation}
As expected, the scale of the low momentum structure is $a/\lambda
^{2}$. However, one should note
that the large value of the numerical factor
implies that the
range of validity of the calculation is limited to very small values
of $a/\lambda$ (so that the
condition $k_c\ll \lambda^{-1}$ is fulfilled).

The energy spectrum obtained with this analytical model is only
self-consistent for wavevectors $k \ll k_c$. In the limit $k \gg k_c$
we assume that
$\varepsilon_k$ goes over to the free particle spectrum
$\varepsilon_{k}\simeq \hbar^2 k^2/2m$, ignoring here a logarithmic
correction (see
Sec.~V).  We smoothly interpolate between these limits, writing
\begin{equation}
U(k)=\frac{2m}{\hbar^2} \left( \Sigma(k,0)-\Sigma(0,0) \right) =
\frac{k_{c}^{1/2}k^{3/2}}{
1+(k/k_{c})^{3/2} }.
\label{interpolate}
\end{equation}
Thus
we estimate the critical temperature as
\begin{equation}
\frac{\Delta T_{c}}{T_{c}}=
\frac{4 \lambda}{3 \pi \zeta(3/2)} \int_0^{\infty} \frac{U(k)}{k^2 +U(k)}
\simeq 3\, an^{1/3}.  \label{f7}
\end{equation}
While the precise coefficient is sensitive to the details of the interpolation
between the low and high $k$ limits, e.g., (\ref{interpolate}), the result remains 
of order unity in any case.  

The $k^{3/2}$ spectrum is only an approximation and is not stable if
higher order corrections are included; from the
general theory of phase transitions, at $T_{c}$, $\varepsilon_{k}\sim
k^{2-\eta}$, where $\eta =\varepsilon^2/54\simeq 0.02$ in an
$\varepsilon=4-D$ expansion or $\eta=8/(3 \pi^2N) \simeq 0.14$ in
the large $N$ limit (with $N=2$) \cite{Zinn}. This model
provides  too strong a modification of the spectrum at small momenta.

The model calculation  illustrates however the basic mechanism behind
the change of
the degeneracy parameter, the modification of the single particle
energy spectrum at small momentum. This may be understood as a result of
correlations among particles caused by their repulsive interactions:  particles
minimize their repulsion by avoiding each other in space, i.e., by
correlating their positions; the physical origin of the effect is therefore a
spatial rearrangement that affects the atoms with low momentum.  By
contrast, atoms with high  momenta  have too much kinetic energy to
develop significant correlations.  The  modification of the spectrum
translates into a
modification of the population of the various levels. In particular
the low momentum
levels at momentum scale
$\sim k_c$ are less populated than they would in a mean field
approximation at the same
density, and the overall result is a decrease of the critical density.

The hardening of the spectrum obtained as a solution of the
     self-consistent equations, which is responsible for the decrease of
the critical
density, also provides a cure for the infrared divergences which occur in the
perturbative calculation in second order. However, as we shall see in
the next section,
such divergences appear in all orders in perturbation theory, so that we
need a more general
scheme to approach the problem.

\section{Classical field approximation}

\label{classical field}

In this section we extend the discussion of the previous section in a way
that is at the same time more general, in that it is not restricted to any
particular class of diagrams, and less general, in that only the linear corrections
to the density are investigated. 

\subsection{Breakdown of perturbation theory}

\label{zero}

Our main goal is the calculation of the critical density. As an intermediate
step, we distinguish in Eq.~(\ref{g1}) for
$\rho _{k}$ the contribution of zero and non-zero Matsubara frequencies:
\begin{equation}
\rho _{k}=-T G(k,0)-T \sum_{\nu \neq 0}\,G(k,i\omega _{\nu }).  \label{g6}
\end{equation}
The density is obtained by integrating over momentum $k$ (see
Eq.~(\ref{densityrhok})). The terms with $\nu \neq
0$  are regular at small momentum since a non-vanishing Matsubara
frequency provides an
infrared cutoff. They provide  corrections to the density, that are
analytic in the self-energy, and therefore of the same order as $\Sigma$, 
starting at order $a^2$ (modulo possible logarithmic corrections). On
the other hand,
     the integral for $\nu =0$ is  singular for small $k$, and the infrared
divergences introduce non-analyticity in $a$. Since, we are
interested here in
the dominant  correction to the critical density, we will retain only
this term in $\rho_k$. Note that the resulting expression for the
density is ultraviolet
divergent, a problem  bypassed by calculating the change in the critical
density.

As illustrated by the example of the previous section, infrared
divergences also occur in the calculation of self-energies $\Sigma
(k,0)$; we 
now use simple power counting arguments to analyze these divergences.  Let
us first consider
     diagrams in which all the internal lines carry zero Matsubara
frequencies. It is
convenient here to introduce a new notation and set
\begin{equation}
\varepsilon^0_{k}-\mu ^{\prime }=(k^{2}+\zeta ^{-2})/2m.  \label{epsilon}
\end{equation}
The quantity $\zeta $, a the mean field correlation
length, is given by
\begin{equation}
\frac{\hbar ^{2}}{2m\zeta ^{2}}=-\mu'=-(\mu -\Sigma _{HF});  \label{zeta}
\end{equation}
$\zeta$ plays the role of an infrared cutoff in the integrals. Note that $\zeta
\rightarrow \infty $ ($\mu ^{\prime }\rightarrow 0$) when
$T\rightarrow T_{c}^{0}$.
In the perturbation series,
we take the intermediate propagators to be
neither free, nor fully self-consistent as
in the  previous
section, but containing the  mean field
contributions. All the
functions that are integrated in the diagrams then appear as products
of fractions of the
form
\begin{equation}
\left[ K^{2}+\zeta ^{-2}\right] ^{-1},
\label{sum}
\end{equation}
where $K$ denotes a generic combination of momenta; it is then
natural to use the dimensionless products $K\zeta $ as new
integration variables.
Consider then a diagram of order $a^{n}$. The lowest order $n=2$ has
been already
explicitly written in (\ref {g10}), and  it is proportional to $ (a/\lambda
)^{2}\ln(k\zeta)$, where $k$ is the external momentum.
For $n>2$, every additional order brings in one factor $a$ from the
vertex,  one
integration over three-momenta, a factor $T$, and two Green's
functions (the internal
lines).
The contribution of the diagram can thus be written as:
\begin{equation}\label{series}
T\,\left(\frac{a}{\lambda }\right)^{2}\left(\frac{a\zeta }{\lambda
^{2}}\right)^{n-2}F(k\zeta ),
\label{cl8bis}
\end{equation}
where $F$ is a dimensionless function, which we do not explicitly need 
here. The main point is that when one approaches the
critical temperature, the
coherence length becomes large so that the summation of terms (\ref{series})
diverges. In the
critical region, $\zeta$ is $\sim\lambda^2/a$, so that
all the terms in
the perturbative expansion are of the same order of magnitude. Therefore, at the
critical point, perturbation theory is not valid.

Let us now assume that in a given diagram some 
propagators carry
non-zero Matsubara frequencies so that one momentum integration will
be altered. For that
integration, the presence of an additional imaginary term
$2i\pi \nu T $ in the denominators of the propagators ensures that no
singularity at $k=0$
can take place. Essentially, in the corresponding
propagators, $\zeta$ is
replaced by a term proportional to $\lambda$, so that  one factor
$a\zeta /\lambda ^{2}$ in (\ref{cl8bis}) is now replaced by $
a/\lambda $. Compared 
to the diagram with only vanishing Matsubara frequencies,
this diagram is down by
a factor $a/\lambda$, and thus negligible in a leading order calculation
of $\Sigma$.

\subsection{Classical field approximation}

The diagrams where all Matsubara frequencies vanish are those of an
effective theory for static fields. Ignoring the non-zero Matsubara
frequencies is indeed equivalent to ignoring the (imaginary) time
dependence of the field
operators. In this approximation the many-body problem reduces to a
classical field
theory in three space dimensions.

The energy of a classical
field configuration
is given by 
\begin{equation}\label{classicalenergy}
{\cal H}= \int d^3r \left( \frac{|\nabla \varphi^(r)|^2}{2m}
-\mu' |\varphi(r)|^2
+\frac{2\pi a}{m}|\varphi(r)|^4 \right).
\end{equation}
The zero Matsubara component of the density is given by
$\langle |\varphi(r)|^2 \rangle$.
By assumption, the wavenumbers of the classical field are limited to
$k$ less than an ultraviolet cutoff
$\Lambda \sim \lambda^{-1}$. As one approaches the critical region,
$k \simle k_c$,
all the terms in the integrand of
(\ref{classicalenergy}) become of the same order of magnitude:
\begin{equation}
\frac{k_c^2}{2m}\sim \mu'\sim \frac{a}{m} \frac{T}{\mu'} k_c^3,
\label{ppp}
\end{equation}
where $(T/\mu')k_c^3$ is the contribution to the density of the modes
with $k\sim
k_c$. From Eq.~(\ref{ppp}) we see that $k_c\sim a/\lambda^2$.  For $k\simeq k_c$
perturbation
theory in
$a$ makes no sense, and in fact all terms in the perturbative
expansion are  infrared
divergent. For
$k_c\ll k\ll
\lambda^{-1}$, perturbation theory is applicable. Note that, in the critical
region, $\zeta\sim 1/k_c\sim
\lambda^2/a$.

By a simple rescaling of the fields $\varphi\to\sqrt{mT}\phi$, one can
write the effective
action for the classical field theory as
\beq
-{\cal H}/T=\int d^3r
\left(\frac{1}{2}|\nabla\phi(r)|^2-m\mu'
|\phi(r)|^2+\frac{4\pi^2 a}{\lambda^2}|\phi(r)|^4\right).
\eeq
The rescaled fields $\phi$ have the dimensions of an inverse length.
The  classical theory  contains ultraviolet
divergences, which spoil simple dimensional
arguments for the linear change of $T_c$.

\subsection{Linear dependence of the density correction}

\label{class}

We now consider a
diagrammatic expansion of
$\Sigma $ in terms of the full zero frequency Green's function, defined by:
\begin{equation}
-2mG^{-1}(k)=k^{2}-2m\mu +2m\Sigma (k,a,G,\Lambda ),  \label{cl4}
\end{equation}
from here on we obmit the explicit index $\nu=0$ in $\Sigma$
and $G$.
In this self consistent expression the self consistent
the self-energy $\Sigma$
depense on $\mu$ only through
its dependence on $G$. Instead of
$\mu $, we use the dimensionless parameter $\alpha $
defined by
\begin{equation}
-2m\mu +2m\Sigma (0)=\alpha \frac{a^{2}}{\lambda ^{4}}.  \label{cl15}
\end{equation}
The parameter $\alpha $ controls the distance to the critical point; it
vanishes exactly at the transition, as opposed to $\mu$.
In terms of $\alpha$
the Green's function is now given by:
\begin{equation}
-2mG^{-1}(k)=k^{2}+\alpha \frac{a^{2}}{\lambda ^{4}}+U(k),  \label{cl15bis}
\end{equation}
where
\begin{equation}
U(k)=2m\Sigma (k,a,G,\Lambda )-2m\Sigma (0,a,G,\Lambda ).  \label{cl15ter}
\end{equation}
Since $\Sigma $ depends only on the full Green's function, $U(k)$ depends
only on $\alpha $ and not on $\Sigma (0)$; moreover, the
ultraviolet divergence in $\Sigma (k)$ is only logarithmic, and the
difference $U(k)$ is independent of 
the cutoff $\Lambda$ 
in the limit $\Lambda\to\infty$.

If we assume that
$\Lambda\to\infty$, the power
counting analysis of
\S
\ref{zero} implies that
\begin{equation}
U(k)=\frac{a^{2}}{\lambda ^{4}}\,\widetilde{\sigma} \left(\frac{k\lambda
^{2}}{a},\alpha \right).
\label{cl-4}
\end{equation}
Inserting this result into (\ref{u13}) and making the change of
variable $x=k\lambda^2/a$, one finds
\begin{equation}
\Delta n_{c}=-\frac{2a}{\pi \lambda ^{4}}\lim_{\alpha \rightarrow 0}\int dx
\frac{\widetilde{\sigma} (x,\alpha )}{x^{2}+\widetilde{\sigma} (x,\alpha )},  \label{cl-6}
\end{equation}
showing that the change in the critical density is indeed linear in $a$.

This result assumes that the limit
$\widetilde{\sigma}(x,\alpha\to 0)$ is well defined. This is 
the case in the self-consistent schemes that we discussed above; they 
avoid the infrared problem of perturbative calculations,
and  lead to well defined values of $\Delta
n_c$. Similarly, in
calculations involving resummations of bubbles or ladder diagrams
the cutoff is provided by
an effective screening explicitly generated by the infinite
resummations. Large $N$ techniques lead to a
similar screening, with the advantage of also providing an expansion
parameter \cite{BigN,BigN2}.

On the other hand, situations where the limit $\widetilde{\sigma}(x,\alpha\to 0)$
is problematic are encountered in perturbation theory where, for reasons
discussed above, an infrared cutoff is needed; 
determination of this cutoff through the condensation
condition can lead to 
spurious $a$ dependence (an explicit example is worked out in detail
in the next section).

The linearity of  the shift in the critical density does not
depend on the ultraviolet cutoff and is thus an universal quantity.
Nevertheless, the universal behavior 
implicity assumes that the limit
$\Lambda \to \infty$ has been taken and is strictly valid only in the limit
   $a\to 0$. If $a$ is not sufficiently small, the classical field
approximation ceases to be valid and non-linear corrections appear. 
The classical field approximation requires that all momenta
involved in the various integrations  are small in comparison
with  $\Lambda \sim \lambda^{-1}$  or, in other words, that the integrands
are  negligibly small for momenta
$k\sim \lambda^{-1}$. Only then, for instance, can we use the approximate
form of the statistical
factors (\ref{u10}). This requires in particular that $k \sim k_c \ll
\Lambda$, yielding
$a/\lambda \ll 1$. In fact, because, as we shall see,
the relation between $k_c$ and
$a/\lambda$ involves a large
number, this regime is reached only for very small
values, $a/\lambda\simle
10^{-2}-10^{-3}$. 

\section{Explicit calculations}

\label{numerical}

Our goal in this section is to provide specific illustrations of
the discussions of the
previous sections. We first present analytical calculations which
shed light on the difficulties
encountered when attempting to calculate the shift in the critical
temperature using
perturbation theory. Then we show how partial resummations of the
perturbative expansion generate
    screening of long range correlations and allow an explicit
calculation of the self-energy, and
then of the transition temperature. Finally we  present results of numerical
self-consistent calculations, which we compare with the analytical
counterparts, and evaluate the limitations of the classical field theory. The accuracy of such approximative
schemes is difficult to gauge a priori. An
alternative is to use
lattice calculations to solve
the three-dimensional
classical field theory. Results of such calculations have been presented
recently
\cite{svistonov,arnold}.

\subsection{Second order perturbation theory}

\label{example}

In order to illustrate the difficulties that one meets in perturbative
calculation near $T_{c}$, let us return for a moment to the second order
self-energy diagram, which is the lowest order diagram that introduces
correlations and therefore corrections to the critical density.
The value of this diagram for vanishing Matsubara frequencies
is given by Eq.~(\ref{g10})
\begin{equation}
\Sigma (k)-\Sigma (0)=
-2 g^2 T^2 \int \frac{d^3 k'}{(2\pi)^3}
\frac{d^3 q}{(2\pi)^3} \frac{1}{(\varepsilon^0_{k'}
-\mu') (\varepsilon^0_{k'-q}-\mu')}
\left[ \frac{1}{\varepsilon^0_{k+q}-\mu')}
- \frac{1}{\varepsilon^0_{q}-\mu'} \right],
\label{sigsumnew}
\end{equation}
where $g=4\pi a/m$.

We note that $\Sigma(k)$ is the convolution of three factors
of the form $1/(k^2+\zeta^{-2})$, with $\zeta$ defined in
Eq.~(\ref{zeta}). Using the
Fourier transform
\begin{equation}
\int \frac{d^{3}k}{(2\pi )^{3}}\,\frac{\text{{\rm e}}^{i k  r}
}{k^{2}+\zeta ^{-2}}=\frac{1}{4\pi r}e^{-r/\zeta },  \label{epsilonr}
\end{equation}
we obtain
\begin{equation}
2m\Sigma (k)=-128\pi ^{2}\left( \frac{a}{\lambda ^{2}}\right) ^{2}\int
r^{2}drj_{0}(kr)\left( \frac{e^{-r/\zeta }}{r}\right) ^{3},  \label{massop}
\end{equation}
where $j_{0}(x)\equiv \sin x/x$. This expression contains, as
anticipated, a
logarithmic divergence at small distances. Let us isolate this divergence by
separating the Bessel function $j_{0}(x)$ into its value at the origin and a
correction term:
\begin{equation}
j_{0}(x)=j_{0}(0)+\left( \frac{\sin x}{x}-1\right)  \label{jzero}
\end{equation}
The first term gives a momentum independent contribution, given by $\Sigma(0)$.
Introducing a cutoff $1/\Lambda $ to
control the ultraviolet
divergence, we obtain
\beq
2m\Sigma (0)=128\pi ^{2}\left( \frac{a}{\lambda ^{2}}\right) ^{2}\mathrm{Ei}(-
\frac{3}{\Lambda \zeta })\approx -128\pi ^{2}\left( \frac{a}{\lambda
^{2}}\right)
^{2}\left[ \ln (\frac{\Lambda \zeta }{3})-\gamma \right],  \label{sigmadiv}
\eeq
where $\gamma = 0.577 \ldots$ is Euler's constant, and $\mathrm{Ei}$ the exponential
integral function. The last approximate equality is valid when
$\Lambda\zeta\gg
1$, which we assume to be  the case. The second term, which is
regular and equal to
$2m(\Sigma ( k)-\Sigma (0))=U(k)$, does not require a cutoff. The result is
\beq
U( k)&=&-128\pi ^{2}\left( \frac{a}{
\lambda ^{2}}\right) ^{2}\int_{0}^{\infty }(j_{0}(kr)-1)\,\frac{e^{-3r/\zeta
}}{r}\,dr\,\nonumber\\  &=&128\pi ^{2}\left( \frac{a}{\lambda ^{2}}\right)
^{2}\left\{ \frac{3}{ k\zeta }{\arctan}\frac{k\zeta }{3}+\frac{1}{2}\ln
\left( 1+(\frac{k\zeta }{3} )^{2} \right)-1 \right\} .  \label{finitepot}
\eeq
This equation implies that  $U(k)$ is a monotonically
increasing function of
$k$, $\sim k^2$ at small $k$, and  growing logarithmically at large $k$.
This logarithmic behavior, obtained in perturbation theory, remains
in general the
dominant behavior of
$U(k)$ at large $k$,
i.e., for
$\zeta^{-1}\ll   k \, \la \, \Lambda$.

Our result for $U(k)$ can now be used in Eq.~(\ref{u13}) in
order to determine the
change in the critical density from Eq.~(\ref{u13}).
Because  $U(k)>0$, this change
is negative. We get:
\beq
\Delta n_c
=-\frac{2}{\pi\lambda^2}\int_0^\infty
{\rm d}k
\frac{U(k)}{k^2+U(k)}
     = -\frac{2}{\pi\lambda^2}\, k_c\, \int_0^\infty {\rm
d}x\frac{J\sigma(x)}{x^2+J^2\sigma(x)},
\eeq
where we have set
\beq
     x=k\zeta\qquad\qquad k_c = 8\pi^2 \frac{a}{\lambda^2}, \quad \quad
U(k)\equiv k_c^2 \,\sigma(x),\qquad\qquad J =\zeta k_c.
\eeq
Note that for small $x$, $\sigma(x)\sim x^2/27\pi^2$  while at large $x$,
$\sigma(x)\sim 2\,(\ln (x/3)-1)/pi^2$. The function
$\sigma(x)/x^2\propto \Delta
\varepsilon_k/\varepsilon_k$ gives an indication, independent of the specific
values of the parameters, of the range of values of
$x$ over which the single particle spectrum is significantly
modified, and hence of
the range of momenta contributing to $\Delta n_c$: this function 
monotonically decreases with $x$,
reaching half its maximum value for
$x\approx 7$, and about $1/10$ of its maximum  when $x\approx
20-30$. Comparison of this momentum scale with the characteristic momentum
scale of higher Matsubara frequencies ($1/\lambda \sim 1/\Lambda$) 
gives a constraint on the
values of $a$ for which the calculation is meaningful. In particular, 
$a$ has to be
small enough  that $a/\lambda \ll \frac{1}{30} ( J/8\pi^2 )$.

As noted earlier, the momentum dependence  of $\Sigma(k)$
is essential for
$\Delta n_c$ to be non-vanishing.
In this second order calculation, the (statistical quasiparticle)
spectrum remains
quadratic at small
$k$, and is given by
\beq
\epsilon_k \to \frac{k^2}{2m} \left(  1+\frac{J^2}{27\pi^2}\,  \right),
\quad \quad \mathrm{for} \quad k \to 0.
\eeq
As expected, the spectrum of the interacting system is harder
than the free spectrum.
In the present approximation, it is identical to the spectrum of 
non-interacting particles
with  an effective mass
$m^*<m$.

     The final result for $\Delta n_c$ depends on the infrared cutoff
$\zeta$, which can be determined by the
condensation condition
$\mu'=\Sigma(0)$:
\beq\label{zetacondens}
\frac{1}{\zeta^2}=\frac{2}{\pi^2}\, k_c^2\,
\left[  \ln\left( \frac{\Lambda\zeta}{3} \right) -\gamma  \right].
\eeq
In principle the ultraviolet cutoff $\Lambda$ could be eliminated by
an appropriate
counter term calculable in the full theory. Alternatively, one could calculate
$\Sigma(0)$ from the expression (\ref{g7}) involving the complete statistical
factors. The result
of such a calculation would be to replace  the term
$\ln(\Lambda\zeta)$  in Eq.~(\ref{sigmadiv}) by
$\ln(\zeta/\lambda)$, up to a numerical additive constant.  Here
we shall simply
choose an ultraviolet cutoff
$\Lambda=1/\lambda$, keeping in mind that there is arbitrariness in the
procedure which affects the final result, since for
$a/\lambda\ll1$, then $\zeta\gg \lambda$, and
$\ln(\zeta/\lambda)$ will
eventually  dominate. Nevertheless, it is intructive
to solve the equation above for
$\zeta$ as a function of $a$ with this choice of cutoff.
Typical values are given in the table below:
\vspace*{.5cm}
\begin{center}
\begin{tabular}{|r|r|r|r|}\hline
$a/\lambda$ &$\zeta/\lambda$ & $J$ & $c$ \\ \hline
0.01     & 6.5     &5.1   &2.4  \\ \hline
0.001    & 23     & 1.8 &0.90  \\ \hline
$10^{-4}$   & 153    &1.2   &0.60  \\ \hline
$10^{-5}$  & 1208    & 0.95  &0.47 \\ \hline
$\cdots$ & $\cdots$ & $\cdots$ & $\cdots$ \\ \hline
$10^{-9}$  &$ 7.5 10^6$   &0.59   &0.29 \\ \hline
$10^{-10}$ &$ 7.0 10^7$   &0.55   &0.27 \\ \hline
$10^{-11}$ &$ 6.5 10^8$   &0.51     &0.25 \\ \hline
\end{tabular}
\end{center}
\vspace*{.5cm}
The behavior of $\zeta$ with $a$ is understandable: if $a$ is large,
condensation
takes place far from the mean field value, hence the small $\zeta$. If $a$ is
small, condensation takes place near the mean field value for which $\zeta\to
\infty$. In fact Eq.~(\ref{zetacondens}) shows that $\zeta\sim
\lambda^2/a$ up to a
logarithmic correction.  The last column of the table gives the
coefficient $c$ in
Eq.~(\ref{deltaTc}) for
$\Delta T_c$.
The variation of $\Delta T_c$ with $a$ follows closely that of
$J\int dx \,\sigma(x)/x^2=J/6\pi$;
that is, the term in $J^2$ in the denominator plays almost
no role.\footnote{Note that these second order results
are closely related to the pertubative calculation of \cite{HGL}
where $c \sim 1$ was obtained by looking at values $0.001 \le
a/\lambda \le 0.01$.}

This simple calculation  also illustrates the
limits of a pertubative approach. The infrared cutoff introduces a new
scale in the problem which spoils the argument leading to the linearity of
the $a$-dependence of $\Delta T_c$ ($c$ is not a constant).  The
condensation condition (\ref{zetacondens}) which relates the infrared cutoff
to the  microscopic length $\lambda$, induces a spurious logarithmic
correction which does not vanish as
$a\to 0$.

\subsection{Non self-consistent bubble sums}

The previous calculation illustrates
how the mixing of ultraviolet and infrared divergences in
perturbation theory can produce spurious
$a$ dependences.
It is therefore desirable to
find approximations in
which the infrared cutoff is internally generated. One such
approximation was already
presented in Sec.~IIIC. We turn now to another, the
resummation of 
bubble diagrams, as illustrated in Fig.~\ref{figgreen1}b.
Again, the
quality of
such an approximation
can only be gauged by a comparison with
an exact calculation, except in large $N$ limit
where the bubble summation becomes exact itself \cite{BigN,BigN2}.

The one bubble diagram
can be calculated explicitly. Keeping an infrared cutoff, we have
\beq
B(q)=T\int\frac{d^3 p}{(2\pi)^3}
\frac{1}{(\varepsilon_p^0 -1/2m\zeta^2)(\varepsilon_{p+q}^0-1/2m \zeta^2 )}
=\frac{4\pi}{\lambda^4
Tq}\arctan \frac{q \zeta}{2}.
\eeq
In the infinite cutoff limit ($\zeta\to\infty$) this simplifies into
\beq
B(q)=\frac{2 \pi^2}{\lambda^4 T}\frac{1}{q}.
\eeq
The Fourier transform of $B(q)$ is nothing but the leading
contribution to the density-density correlation function. At the
critical point this correlations
behaves as $1/r^2$, so that density fluctuations are correlated
over very large
distances; this is the physical origin of the infrared divergences of
perturbative
calculations. Nevertheless, these fluctuations can be screened, for instance 
by summing the bubble
or ladder diagrams. The respective contributions of the two classes of
diagrams  actually differ only  by the number of exchange diagrams. For the
bubble sum, the
correlation  function reads
\beq\label{bubblessum}
\frac{ B(q)}{1+ 2 g B(q)}=\frac{2\pi^2}{\lambda^4 T}
\frac{1}{q+k_c},
\eeq
and is now regular at small $q$.
   The screening wave number $k_c$ is given by
\beq
k_c=8 \pi^2 \frac{a}{\lambda^2}.
\label{kkkc}
\eeq
In the ladder approximation the factor of $2$ in the denominator of
(\ref{bubblessum}) is absent, and, correspondingly, $k_c=4\pi^2 a/\lambda^2$.

Now, an infrared cutoff
is no longer needed in the calculation of $U(k)$, and the limit
$\zeta\to\infty$ can be taken. One finds
\beq
U(k)=-\frac{2}{\pi^2 } k_c^2 \,\kappa\,\int_0^\infty d x
\,\frac{1}{1+x \kappa}\,\left[\frac{x}{2}\,\ln\left|
\frac{1+x}{1-x}\right|-1\right],
\eeq
where $\kappa\equiv k/k_c$.
To study the limiting behavior of $U(k)$ at large and small $k$, it is
convenient to transform this expression as follows. First we integrate twice
by parts to obtain
\beq
U(k)=-\frac{2}{\pi^2 } \frac{k_c^2}{\kappa}
\int_0^{\infty}  d x\,
\frac{2 (1+ x \kappa)[ \ln( 1+ x \kappa) -1]}{(1-x^2)^2}.
\eeq
Taking the derivative of the integrand with respect to $\kappa$
which obtain
\bea
\frac{d}{d \kappa} & &\int_0^{\infty}  d x\,
\frac{2 (1+ x \kappa)[ \ln( 1+ x \kappa) -1]}{(1-x^2)^2} \nonumber \\
& = &\frac{1}{2} \int dx \ln(1+ x \kappa) \left(
\frac{1}{(1-x)^2}-\frac{1}{(1+x)^2} \right)
    =  \frac{\kappa^2}{1-\kappa^2} \ln \kappa.
\label{deriva}
\eeq
The $\kappa$ integral can now be expressed in terms of
the  polylogarithmic function $g_p(x)$, defined in Eq.(\ref{ig2});
the integration constant is choosen to make $U(k=0)=0$. Thus,
\beq
U(k)=-\frac{2}{\pi^2  } \frac{k_c^2}{\kappa}
\left\{ \kappa [1-\ln \kappa] + \frac{1}{2} \ln \kappa \ln (1+\kappa)
+\frac{1}{2} \left[g_2(1-\kappa) + g_2(-\kappa)
\right]-\frac{\pi^2}{12}
\right\}.
\label{horrform}
\eeq
However, to derive the limiting behavior of $U(k)$ it is more convenient to
take the limits in Eq. (\ref{deriva}) and integrate afterwards.

For small $\kappa$ ($\kappa=k/k_c \simle 0.1$), $U(k)$ is well approximated by
its small
$k$ behavior:
\beq
U(k)=-\frac{2}{3 \pi^2 }  k^2 \left( \ln \frac{k}{k_c} - \frac{1}{3}
\right), \quad \quad k\ll k_c.
\eeq
As expected from perturbation theory,
$U(k)$ grows logarithmically for large momenta, $k$,
and for $k/k_c \simge 50$, $U(k)$  is well approximated
by:
\beq
U(k)=\frac{2 k_c^2}{ \pi^2  } \left( \ln \frac{k}{k_c} -1 \right),
\quad \quad k/k_c \gg 1.
\eeq

   From the small
$k$ behavior of
$k^2+U(k)$ one can estimate the critical index $\eta$.
The logarithmic term indicates a modified
power law in the low momentum limit
$\sim k^{2-\eta} \sim k^2 (1-\eta \ln k +...)$.
Comparing the coefficients of the logarithmic terms
we obtain
\beq
\eta=\frac{2}{3\pi^2} \approx 0.068.
\eeq
Due to the exchange contributions this value differs by a factor
of $2$ from the usual large $N$ results.\footnote{Note however that the
expansion in powers of $\eta$ is meaningful only if the magnitude of $\eta$ is
controlled by a small parameter, such as in the $\epsilon$-expansion or the
$1/N$-expansion. The estimate presented here should therefore not be viewed
as a particular prediction for the critical index $\eta$; it gives
nevertheless an
indication of how the spectrum is modified at small $k$ by the resummation
of particle-hole bubbles. Another estimate of the effect of bubble summation
was presented in Ref.~(\cite{3/2club}); there we tried to estimate the change of
the spectrum with respect to the $k^{3/2}$ self-consistent solution.
Once the bubble sum is included however, 
self-consistency does not further alter the spectrum at low momentum, as 
later in this section. As a result, the exponent $\eta$ that one finds
here is much
smaller than the crude estimate in Ref.~(\cite{3/2club}). }

The change in the critical density is now
\beq\label{bub5}
\Delta n_c=-\frac{2 k_c}{\pi\lambda^2}
\int_0^\infty
d \kappa\,\frac{\sigma(\kappa)}{\kappa^2+\sigma(\kappa)}.
\eeq
Let us first estimate the range of $\kappa_0=k_0/k_c$ where
$\sigma(\kappa)$ dominates over $\kappa^2$. 
Using the small
$k$ asymptotics of $U(k)$ we estimate
$\kappa_0 \, \la \, \exp[- 3 \pi^2/2] \ll 1$.
Therefore, we can  again ignore the term in $\sigma(\kappa)$ in the
denominator of
(\ref{bub5}) without making a significant error; it only brings in an harmless
singularity at small $\kappa$. We get then:
\beq
\frac{\Delta n_c}{n_c}=
\frac{4 k_c \lambda}{\pi^3 \zeta(3/2)}
\int_0^\infty \frac{ d
\kappa}{\kappa}\int_0^\infty
             \frac{ d x}{1+x\kappa}
              \left(\frac{x}{2}\ln\left|\frac{1+x}{1-x}\right|
         -1\right).
\eeq
In order to calculate the integral, we want to exchange the
orders of the $\kappa$ and $x$ integrals.  Since the
integrals, however, are not absolutely convergent, before we do so we need to
     introduce a regularization, inserting  a factor
$\kappa^\epsilon$ in the $\kappa$ integral, and taking the limit $\epsilon\to
0^+$.  With this factor we may exchange the orders of integration.
The $\kappa$ integral becomes
\beq
\int_0^\infty d \kappa \,\frac{\kappa^{\epsilon - 1}}{1+ x \kappa} =
\frac{1}{\epsilon  x^\epsilon}.
\eeq
The remaining $x$ integral
becomes
\beq
      \int  d x\, x^{-\epsilon}
              \left(\frac{x}{2}\ln\left|\frac{1+x}{1-x}\right| -1\right).
\eeq
For $\epsilon = 0$ this integral vanishes identically.  Thus we may replace
$x^{-\epsilon}$ by $x^{-\epsilon}-1$ which goes to $-\epsilon \ln x$ as
$\epsilon \to 0$.  The remaining integral is
\beq
      \int d x\, \ln x
          \left(\frac{x}{2}\ln\left|\frac{1+x}{1-x}\right| -1\right) = -\frac
{\pi^2}{8}.
\eeq
The factors of $\epsilon$ cancel out, and we find
\beq
\frac{\Delta n_c}{n_c}=
-\frac{k_c \lambda}{2 \pi \zeta(3/2)}
=- \frac{4 \pi}{\zeta(3/2)}   \frac{a}{\lambda}.
\eeq
We finally obtain the changes in
the transition density and the transition temperature:
\beq
\frac{\Delta n_c}{n_c} =
      - \frac{4\pi}{\zeta(3/2)^{4/3}}  a n^{1/3}\qquad\qquad
\frac{\Delta T_c}{T_c}
=-\frac{2}{3}\frac{\Delta n_c}{n_c}
     \simeq 2.33\,  a n^{1/3},
\label{result}
\eeq
This result for the bubble sum agrees with the leading order
result of the $1/N$
expansion. It is interesting to observe that the leading order $1/N$ result is
independent of $N$.
Since $a N$ is kept constant in the $1/N$ expansion, 
$k_c$ is effectively independent of $N$ ($k_c=2\pi^2 aN/\lambda^2$),
while $U(k)$ is
of order
$1/N$. Therefore, the approximation of neclecting $U(k)$
in the denominator of
Eq.~(\ref{bub5})
is justified 
in the
$1/N$ expansion.

In the bubble sum, we can keep $U(k)$ in the denominator and calculate
the integral in Eq.~(\ref{bub5}) numerically, and find a reduction 
the linear coefficient of the critical temperature shift from $c=2.33$ to
$c=2.20$ for $N=2$.
In this approximation the condensation condition reads 
\beq\label{condesation2}
2m\Sigma(0)=- \frac{2 k_c^2}{\pi^2} \int_0^{\Lambda/k_c} d x
\frac{1}{1+x^2} = \frac{2 k_c^2}{\pi^2} \ln \frac{k_c}{\Lambda+k_c},
\eeq
which gives the mean field correlation length,
\beq
\frac{1}{\zeta^2}=\frac{2k_c^2}{\pi^2}\ln \frac{1}{\lambda k_c}.
\eeq
As before we have taken $\Lambda=1/\lambda$ and  assumed  that $\lambda k_c\ll 1$.
As opposed
to the second order calculation, Eq.~(\ref{zetacondens}),
the condition (\ref{condesation2})
does not mix the infrared and ultraviolet cutoff,
and does not
introduce any spurious $a$ dependence in the final result for the shift in
the critical temparature.

The condensation condition (\ref{condesation2}) is the only place
where the microscopic scale $\lambda$ enters explicitly. However, 
the classical field theory result for $\Delta
T_c$ assumes implicitly that the contributions of momenta $k\sim
\lambda^{-1}$ are vanishingly small. Alternatively, if we were to cut
the integration in (\ref{bub5}) off at $k\sim \Lambda$,
one should find a result
independent of the specific value of $\Lambda\simge \lambda^{-1}$. In fact
we have seen that the  momenta important in the determination of $\Delta
T_c$ are $k\sim k_c$. The validity of the
classical field approximation requires that  $k_c\ll \lambda^{-1}$,
or, since $k_c=8\pi^2 a\lambda^2$,
$a/\lambda\ll 1/8\pi^2$; thus, the linear regime  is attained only
for anomalously  small $a$.
When $a$
is not so small, non-linear corrections $\sim a^2 \ln
(a/\lambda)$ appear, which
tend to decrease the value of $\Delta
T_c$,
as discussed in Ref.~\cite{NEW} (see also below).

In Ref.~\cite{Stoof1}, Stoof examines the
appearance of Bose-Einstein
condensation, calculating 
the shift of the critical density
within a real time formalism. The
approach includes, not only the
mean field contributions, but also
sums of ladder graphs within
the
many-body T-matrix-approximation.
He derives
an analytical formula
for the
modification of the energy spectrum, from which
Stoof
obtains a relative increase of the critical temperature,
$4.66\,\,an^{1/3}$, exactly twice the value of
the large-N calculation \cite{BigN}.
Summing ladders, and
neclecting $U(k)$ in
the denominator
of  Eq.~(\ref{bub5}) we indeed reproduce
this result. 
Evaluating the entire integral numerically, one obtains $c=3.90$.

\subsection{Self-consistent calculations}

We  now solve numerically the self-consistent calculations discussed in
Sec.~IIIC.  We quantitatively compare
three  different approximations for the self-energy, 
the one bubble
approximation, Eq.~(\ref{smallk0}),
\beq
\Sigma(k)-\Sigma(0)
=-2g^2T \int \frac{d^3q}{(2\pi)^3}
B(q)
\left(
\frac{1}{\varepsilon_{k-q}}-\frac{1}{\varepsilon_{q}} \right) ,
\label{loop}
\eeq
and, to compare with previous calculations,
the ladder summation of particle-particle scattering
processes
\begin{equation}
\Sigma(k)-\Sigma(0)=-2g^2T \int \frac{d^3q}{(2\pi)^3}
\frac{B(q)}{1+gB(q)}
\left(
\frac{1}{\varepsilon_{k-q}}-\frac{1}{\varepsilon_{q}}\right).
\label{ff10}
\end{equation}
and, finally, the bubble summation of
particle-hole scattering
processes
\begin{equation}
\Sigma(k)-\Sigma(0)=-2g^2T \int \frac{d^3q}{(2\pi)^3}
\frac{B(q)}{1+2gB(q)}
\left(
\frac{1}{\varepsilon_{k-q}}-\frac{1}{\varepsilon_{q}}\right),
\label{f10}
\end{equation}
The energy spectrum $\varepsilon_k$ in the denominators are
determined
self-consistently using Eq.~(\ref{noname}); $B(q)$ is
given in Eq.~(\ref{bubb0}).

Although the integrals
in Eqs.~(\ref{loop})-(\ref{ff10})
giving the difference of the self-energies,
$U(k)=2m [\Sigma(k)-\Sigma(0)]$,
are convergent,
we introduce a large
momentum cutoff
$\Lambda$ for their numerical evaluation ($U(k\ge \Lambda) \equiv 0$).
Only in the limiting case $\Lambda \to \infty$, will $U(k)$
become independent of $\Lambda$; for any finite cutoff,
the energy spectrum  depends weakly on $\Lambda$.
The cutoff enters only through the dimensionless
parameter $\Lambda \lambda^2/a$. For the numerical calculation
the value $\Lambda \lambda^2/a \simeq 800$
was used.  For the self-consistent bubble calculation we further studied
the influence of the cutoff to extrapolate numerically to the limit
$\Lambda \to \infty$.

\begin{figure}
\begin{center}
\epsfig{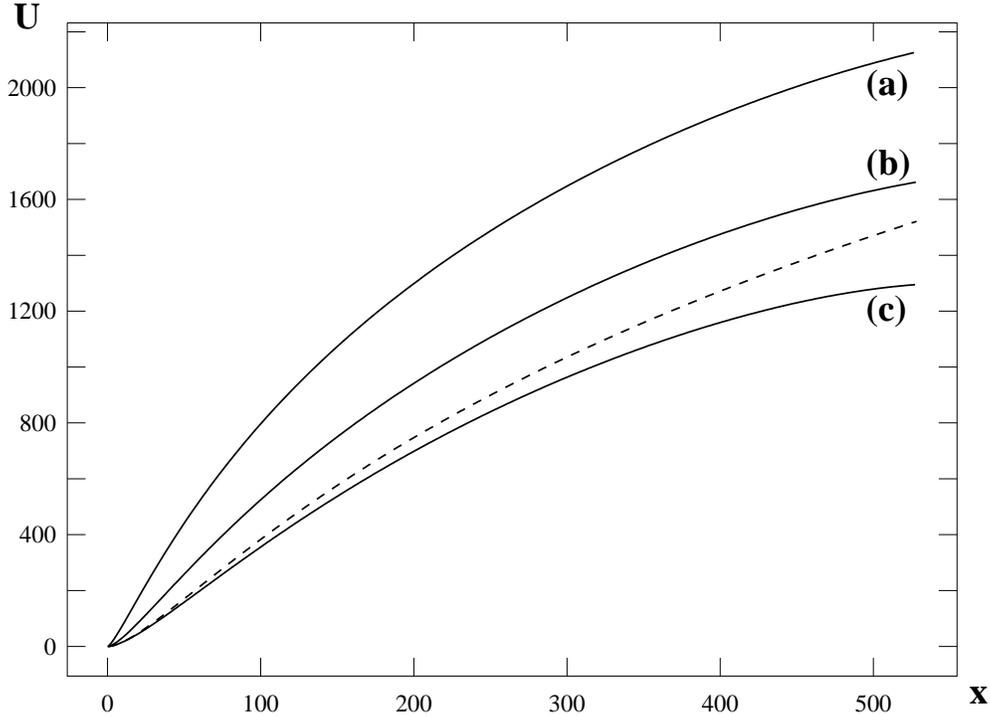}
\end{center}
\caption{\label{figuk11}Self-energy $U(k)$ in units of
$a^2/\lambda^4$ plotted as a function
of $x=k \lambda^2/a$ for the three approximations discussed in the text:
(a) self-consistent one bubble approximation, (b) self-consistent
ladder summation, and (c) self-consistent bubble sum.
These result are obtained
with an ultraviolet cutoff of
$\Lambda \lambda^2/a \simeq 800$.
The dashed line shows the analytical (not self-consistent) calculation of
$U(k)$ in the bubble approximation, Eq.~(\ref{horrform}).
}
\end{figure}

Figure \ref{figuk11} summarizes the numerical results of this section
in terms of the self-energies
$\Sigma(k)-\Sigma(0)$
corresponding to the three different approximations.
The various curves in Fig.~\ref{figuk11} display
the logarithmic
growth of $U(k)$ at large $k$. Note however that within the present
approximations the overall
magnitude of $U(k)$ is
determined by the behavior of the  spectrum at small $k$: the harder
the spectrum,
the larger $U(k)$, and the larger
the value of $c$.
Nevertheless, the values of the shifts in the critical
temperature remain comparable.
For instance, for the  value of the cutoff given above,
the shifts of the critical temperature that we obtain
from Eq.~(\ref{u13})
are:
$\Delta T_c/T_c^0 \simeq 3.8 \,a n^{1/3}$
for the self-consistent one bubble calculation,
$\Delta T_c/T_c^0 \simeq 1.6 \,a n^{1/3}$ for the
self-consistent bubble sum, and
$\Delta T_c/T_c^0 \simeq 2.5 \,a n^{1/3}$ for the self-consistent
ladder sum.
These values still depend weakly on
the value of the ultraviolet cutoff $\Lambda$, and still conatin  logarithmic
corrections $\sim a^2\ln (a/\lambda)$, as we shall see below.

We now compare  in more detail
these results with our
analytical calculations and discuss briefly the extrapolation
$\Lambda \to \infty$, i.e. the extrapolated result 
of the bubble summation is $\Delta T_c/T_c \simeq 2.0 a n^{1/3}$.

\subsubsection{Self-consistent one-bubble calculation}

In the limit $k\rightarrow 0$,
we expect to recover the $k^{3/2}$ behavior of the analytical model
of Sec.~IIIC.
By
fitting the numerical data to the following functional form
\beq
2m(\Sigma(k)-\Sigma(0)) = \frac{k_c^{1/2} k^{3/2}}{1
+ a_{1/2} (k/k_c)^{1/2}
+a_1 (k/k_c) + a_{3/2} (k/k_c)^{3/2} + ...},
\eeq
we extract a momentum scale
$k_c$, which agrees quantitativly with that of the  analytical
calculation, $k_c
\sim 20\, a/\lambda^2$. However, the spectrum very soon deviates from
this behavior, due to
the large value of the coefficient $a_{1/2}
\simeq 0.9$. At intermediate wavevectors, around $k_c$,
$U(k)$ is roughly  linear, and eventually grows
logarithmically for $k \gg k_c$, as expected from pertubation theory.

\subsubsection{Self-consistent bubble sum}

As we have seen in a previous example the main effect of 
self-consistency is
to modify the spectrum at low momentum, avoiding infrared divergencies.
Since, however, the bubble sum already provides a
screening of the long
range correlations leading to the infrared divergences, we do not
expect qualitative
changes in
$U(k)$ in going from the non self-consistent
result of Sec.~VB, to the fully self-consistent calculations.
This behavior 
can be seen in  Fig.~\ref{figuk11}:   deviations occur only at
high momenta,
mainly due to the influence of the finite cutoff in the numerical solution.

To study more quantitatively the influence of a large but finite
cutoff $\Lambda$  we
have  performed a self-consistent calculation of $U(k)$ numerically
for several values of $\Lambda \lambda^2 /a$. As explained in
Ref.~\cite{NEW} we expect a logarithmic dependence on $\Lambda\lambda^2/a$;
therefore we have used this functional form to fit
our numerical data, which provides
\beq
\frac{\Delta T_c}{T_c} \simeq 1.95 \, an^{1/3} \left[ 1 + 32 \frac{an^{1/3}
}{\Lambda\lambda}
     \ln \left( 21 \frac{an^{1/3}}{\Lambda \lambda} \right) \right],
\quad \quad \Lambda/a  \gg 1.
\label{funcform}
\eeq
Extrapolating to $\Lambda \to \infty$ we obtain $c\simeq 2.0$,
which is slightly smaller than the shift obtained for the non self
consistent bubble sum. Alternatevely, taking a finite value for
$\Lambda \lambda$ that is independent of $a$, e.g. $\Lambda \lambda \sim 1$,
provides a logarithmic correction which limits the linear regime
to very small values of $an^{1/3}$.
The precise value of this correction is model dependent,
as we see in the following subsection.

\subsubsection{Influence of non-zero Matsubara frequencies}

Non-linear corrections to the critical
temperature shift cannot  be obtained within the
zero Matsubara frequency sector. One possibility would be
to use an effective field theory which includes
the effects of non-zero frequencies, e.g., as new vertices in the
effective action.
Here, we use a different approach,
solving
the
following pair of non-linear equations,
\beq
\epsilon_k=
\frac{\hbar^2 k^2}{2m} +\Sigma(k)-\Sigma(0),
\quad
\Sigma(k)-\Sigma(0) = -2 g^2 T \int \, \frac{d^3q}{(2\pi)^3}
\frac{B(q)}{1+2 g B(q)}
\left(\frac{1}{\epsilon_{k-q}}-\frac{1}{\epsilon_{q}}
\right),
\label{nonlin1}
\eeq
where the cutoff in the bubble diagram integral
\beq
B(q)\simeq \beta \int  \frac{d^3p}{(2\pi)^3}
f_p f_{p+q}
\label{nonlin2}
\eeq
is no longer a simple step function, but rather the smoother
Bose function $f_k=(\exp(\beta \epsilon_k)-1)^{-1}$. We calculate 
the shift in the transition temperature using
Eq.~(\ref{firstl}).  Although this does not correspond to a systematic
approximation, it provides an illustration of the effect of keeping the
full statistical factors in the calculation (instead of using their
classical limit).
\begin{figure}
\begin{center}
\epsfig{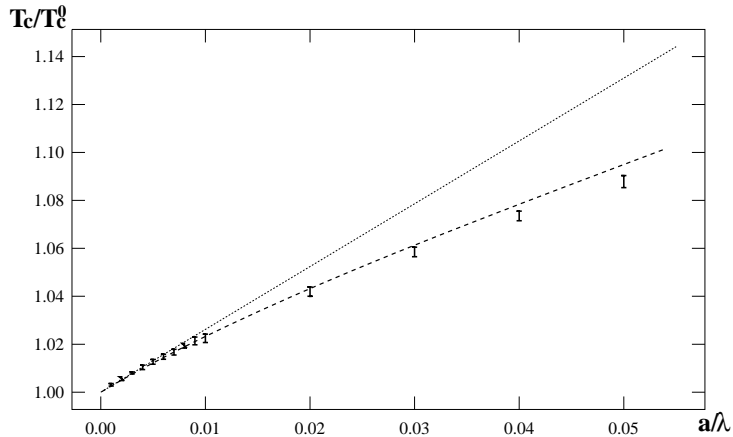}
\end{center}
\caption{\label{figtc1}Dependence of the transition temperature,
$T_c/T_c^0$, of a
dilute homogeneous
Bose gas on scattering length $a/\lambda$,
calculated by solving Eqs.~(\ref{nonlin1}) and
(\ref{nonlin2}) self-consistently. The dashed line is a fit to the data points,
given by Eq.~(\ref{fitdat}). The asymptotic linear behavior
extracted from this fit is shown for comparision. The linear
behavior is seen only
at very small values of $a/\lambda$.}
\end{figure}
In Fig. \ref{figtc1} we show the calculated critical
temperature in the dilute region.
The Bose functions in (\ref{firstl}) lead to  $a^2
\ln a$ corrections  in $\Delta
T_c/T_c$. On the other hand,
the Bose functions in the bubble diagram, Eq.~(\ref{nonlin2})
lead to less singular corrections. We ignore them here and fit the
the numerical datas to the same functional form of Eq.~(\ref{funcform})
as  in the last
subsection, and find,
\beq
\frac{\Delta T_c}{T_c}\simeq 1.9 a
n^{1/3} \left[
1+ 2.6 a n^{1/3} \ln (3.1 a n^{1/3})
\right].
\label{fitdat}
\eeq
Even in the very dilute region,
$n^{1/3} a
\sim 0.01$, the logarithmic corrections are noticeable and
reduce the temperature shift with respect to the linear
prediction.
This provides a possible explanation for the
discrepancy of the different
Monte Carlo results \cite{HK,svistonov,arnold} and \cite{GCL}; whereas Refs.
\cite{HK,svistonov,arnold} calculated the linear corrections directly
in the limit
$n^{1/3}a \rightarrow 0$, Ref. \cite{GCL} performed several
calculations
in the density regime $10^{-6} < na^3 \la 0.1$ finding
a shift of the
critical density much smaller than expected
from the linear formula of
Refs. \cite{HK,svistonov,arnold}. Although the logarithmic
corrections
tend to  decrease this linear shift, the
approximations
underlying Eqs. (\ref{nonlin1}) and (\ref{nonlin2})
are too crude to allow
quantitative
comparision.

In \cite{Stoof2} Bijlsma
and Stoof, using renormalization group techniques, obtained
an increase of
the critical temperature.
A
peculiar feature of their results is that the dependence of the
critical
temperature on the dimensionless parameter $an^{1/3}$ is
given by an unusual
curve, going as $a \ln a$ in the limit of vanishing interaction
\cite{Stoof3}.
The interpretation of
such an unexpected dependence is not
clear at this stage.

\section{Conclusion}
In this paper, we have studied the effects of
particle interactions and correlations
on the transition temperature for Bose Einstein condensation,
and derived the leading  effects beyond mean field in dilute systems. Our study is general
and not limited to any particular approximation, for instance an arbitrary selection of
class of diagrams in a perturbation expansion. We have shown that the leading term in the
change of the critical density is first order in the scattering length $a$, and can be
derived by solving the corresponding classical field theory.
Estimating analytically the coefficient requires in general uncontrolled
approximations.  Among the
various approximations that we have tried,
our prefered result is the self-consistent
calculation of the sum of bubble diagrams, which gives a coefficient 2.0. However this number
should not be trusted at a $10\%$ level. Compared with the most recent
numerical results of Refs.~\cite{svistonov,arnold}, $c \simeq 1.3$,
our value 
is still acceptable;
the complexity of the mathematical problem
does not permit one to make a  definitive 
prediction  of the prefactor of the linear term from an analytic analysis.

It is remarkable, that, despite this complexity, all approximations
that we have used lead to comparable results: to get the right order of magnitude of 
the critical density or
temperature change a precise  determination of the energy
shift $U(k)$ is not required. The contribution of
this function to $\Delta T_c$ are actually close to ''all or nothing'' for extreme
$k$ values: for small $k$,
the function is larger than the free particle energy and the
corresponding momenta are completely
depopulated, the precise value of $U(k)$ is not relevant; for large
$k$, the free particle
energy dominates and the value of $U(k)$ is also irrelevant.
The important feature of $U(k)$ are the
crossover values at which it is comparable to the free particle
energy spectrum, and the way this
region is crossed by the function.

We have limited ourselves to an homogeneous gas contained in a box, ignoring the  
influence of a possible external potential, for example magnetic traps
and optical lattices. 
In both such systems, the dimensionality can vary continuously from three
to two or smaller, and, therefore, affects the nature of the transition.
We will discuss them in future publications.

\begin{center}
ACKNOWLEDGEMENTS
\end{center}
Author GB would like to
thank the Ecole Normale Sup\'erieure
and the CEA Saclay Center, and GB, FL, and MH the Aspen
Center of Physics
for hospitality in the course of this work.  This research
was facilitated by the Cooperative Agreement between the University of
Illinois at Urbana-Champaign and the Centre National de la Recherche
Scientifique, and supported in part by the NASA Microgravity Research
Division, Fundamental Physics Program and by National Science Foundation Grant
PHY98-00978 and continuation.  Laboratoire Kastler Brossel de l'Ecole Normale Sup\'erieure is
{\it UMR 8552 du CNRS} and {\it associ\'e \`a l'Universit\'e Pierre et Marie
Curie}.

\begin{center}
APPENDIX
\end{center}

Since the
formalism of Ursell operators is less common than that of
Green's
functions, we give in this appendix a few more technical
details concerning
the equations written in \S\ \ref{Ursell op}; this
will allow the interested
reader to make contact with the
calculations of ref. \cite{HGL} more easily.
For instance, the right
side of Eq.~(\ref{u2ter}) can be obtained from
Eqs.~(58),
(53) and (55) of this reference, which
provide:
\begin{equation}
X_k=e^{\beta \Delta \mu }\rho_k,
\label{app1}
\end{equation}
so that (55)
becomes:
\begin{equation}
-\log \left[ 1-\frac{4a}{\lambda }\left(
\frac{\lambda }{2\pi }\right)
^{3}\int d^{3}k\,\,\rho_k
\,\times e^{\beta \Delta {\mu
}}\right].
\label{u3}
\end{equation}
Equation ~(\ref{u2ter}) is then
nothing but the first order term in an
expansion of this result in
powers of $a$.

Similarly, Eq.~(\ref{uu6}) can be obtained as
the lowest order
expansion of a relation obtained from Eqs. (55) and
(81) of \cite{HGL},
\begin{equation}
\beta \,\,\delta \mu _{k}=-\log
\left[ 1+8\left( \frac{a}{\lambda }\right)
^{2}e^{3\beta \Delta {\mu
}}{J}_{2}\right]
\label{u5bis}
\end{equation}
with
\begin{equation}
{J}_{2}=\lambda
^{6}\int \frac{d^{3}k^{^{\prime }}}{\left( 2\pi \right) ^{3}}
\int
\frac{d^{3}q}{\left( 2\pi \right) ^{3}}\,\,\rho_{k'}
\,\rho_{k'-q}\,\rho_{k+q}
\times
e^{\beta \Delta (k,k',q)},
\label{u5ter}
\end{equation}
and
\begin{equation}
\Delta (k,k',q)=
=\delta \mu_{k'}
+\delta \mu_{k'-q}+ \delta \mu_{k+q}.
\label{u6}
\end{equation}
We note 
that we have changed the sign convention of \cite{HGL}
by
introducing a minus sign in the right hand side of (\ref{uu6}); in
this way,
positive $\Delta \mu $ as well as positive $\delta \mu
_{k}$ correspond to
positive corrections to the self-energies. This
convention makes more
straightforward the comparison between $\delta
\mu _{k}$ and the self-energy
$\Sigma (k)$ introduced in the Green's
function formalism.

The exact form of the $\delta \mu $'s is not
important for the discussion of
\S\ \ref{Ursell op}.\ In the context
of mean field, what matters actually is
only the existence of some
$k$ independent form of $\Delta \mu $, and
one could use
expression (\ref{u3}) as well; nevertheless, it would not
improve the
accuracy either, since it , but it is actually just a
consequence of
the simplest approximation used for the self-consistent
equation for
$\rho $. As for correlations effects, the only essential
property is
the momentum conservation rule that appears in (\ref{uu6}) as
well as
in (\ref{u5ter}).

        Higher orders can be readily incorporated into the self-consistent
equation \cite{HGL}; for instance, a summation of
bubble diagrams shown in
Fig.~\ref{ursellrho}d leads to the
generalization of
Eq.~(\ref{uu6}):
\begin{equation}
\beta (\delta \mu_k - \delta \mu_0 )=
-4\frac{a}{\lambda }\lambda^3 \int
\frac{d^3 q}{(2\pi)^3}
\frac{A(q)}{1+A(q)}\left( \rho_{k+q} - \rho_{q} \right)
\label{u9}
\end{equation}
where
\begin{equation}
A(q)=2\frac{a}{\lambda }\lambda^3 \int
\frac{d^3 k^{\prime}}{(2\pi)^3}
\,\rho_{k^{\prime}} \,\rho_{k^{\prime}-q}.
\label{u9bis}
\end{equation}
With the factor $A(q)$ in the denominator the integral of Eq.~(
\ref{u9}) is convergent in the infrared with a free particle spectrum.
Further generalizations are discussed in \cite{HGL}. Numerical solutions of
particular approximations are presented in Sec.~V.
As far as the bubble summation of Eq.~(\ref{u9})
is concerned, we
remark that a summation of
Ursell ladder-like diagrams leads to the
same result without the $2$
in the denominator; for more details, see \cite
{Markus}.

\newpage

\end{document}